\documentclass[journal,twoside,web]{ieeecolor}
\usepackage{generic}
\usepackage{cite}
\usepackage{amsmath,amssymb,amsfonts}
\usepackage{algorithmic}
\usepackage{graphicx}
\usepackage{textcomp}
\usepackage{threeparttable}
\usepackage{colortbl}
\usepackage{color}

\definecolor{gray1}{rgb}{0.95,0.95,0.95}
\definecolor{gray2}{rgb}{0.85,0.85,0.85}

\def\BibTeX{{\rm B\kern-.05em{\sc i\kern-.025em b}\kern-.08em
    T\kern-.1667em\lower.7ex\hbox{E}\kern-.125emX}}
\begin{document}
\title{Fundus Image Quality Assessment and Enhancement: a Systematic Review}
\author{Heng Li, \textit{Senior Member}, \textit{IEEE}, Haojin Li, Mingyang Ou, Xiangyang Yu, Xiaoqing Zhang, Ke Niu, \textit{Senior Member}, \textit{IEEE}, Huazhu Fu, \textit{Senior Member}, \textit{IEEE}, and Jiang Liu \textit{Senior Member}, \textit{IEEE}
\thanks{This work was supported in part by the National Natural Science Foundation of China (62401246, 82272086), National Key R\&D Program of China (2024YFE0198100, 2024YFC2510800), Shenzhen Natural Science Fund (JCYJ20240813095112017), Shenzhen Medical Research Fund (D2402014),
and Agency for Science, Technology and Research (A*STAR) Central Research Fund (“Robust and Trustworthy AI system for Multi-modality Healthcare”).
% (Corresponding Author: Heng Li, lih3@sustech.edu.cn, Jiang Liu, liuj@sustech.edu.cn)
}
% \author{Heng Li, Haojin Li, Mingyang Ou, Xiangyang Yu, Xiaoqing Zhang, Ke Niu, Huazhu Fu, and Jiang Liu
\thanks{Corresponding author: Heng Li, Jiang Liu (email: lih3, liuj@sustech.edu.cn).}
\thanks{H. Li, H. Li, M. Ou, X. Yu, and J. Liu are with the Research Institute of Trustworthy Autonomous Systems, SUSTech, Shenzhen, China, and also with the Department of Computer Science and Engineering, SUSTech, Shenzhen, China (e-mail: liuj@sustech.edu.cn).}
\thanks{X. Zhang is with the Center for High Performance Computing and Shenzhen Key Laboratory of Intelligent Bioinformatics, Shenzhen Institute of Advanced Technology, Chinese Academy of Sciences, Shenzhen, China.}
\thanks{K. Niu is with the Computer School, Beijing Information Science and Technology University, Beijing, China.}
\thanks{H. Fu is with the Institute of High Performance Computing (IHPC), Agency for Science, Technology and Research (A*STAR), Singapore.}
}

\maketitle

\begin{abstract}
As an affordable and convenient eye scan, fundus photography holds the potential for preventing vision impairment, especially in resource-limited regions.
However, fundus image degradation is common under intricate imaging environments, impacting following diagnosis and treatment.
Consequently, image quality assessment (IQA) and enhancement (IQE) are essential for ensuring the clinical value and reliability of fundus images.
While existing reviews offer some overview of this field, a comprehensive analysis of the interplay between IQA and IQE, along with their clinical deployment challenges, is lacking.
This paper addresses this gap by providing a thorough review of fundus IQA and IQE algorithms, research advancements, and practical applications.
We outline the fundamentals of the fundus photography imaging system and the associated interferences, and then systematically summarize the paradigms in fundus IQA and IQE.
Furthermore, we discuss the practical challenges and solutions in deploying IQA and IQE, as well as offer insights into potential future research directions.
\end{abstract}


\section{Introduction}
\label{sec:introduction}
%眼底图像的价值
Fundus photography, an affordable and convenient eye scan, has been successfully employed for ocular disease screening and diagnosis worldwide~\cite{WHO_2023}. 
Furthermore, cooperating with intelligent diagnostic systems~\cite{li2023artificial}, fundus photography holds great promise for broadening eye screening in undeveloped regions~\cite{liu2023economic}, significantly contributing to the prevention of blinding eye diseases~\cite{WHO_2022}.
Despite the advantages, fundus image quality often degrades in complex conditions, leading to diagnostic uncertainty~\cite{maier2022image}. 
\begin{figure}[tbp]
    \begin{centering}
        \includegraphics[width=\linewidth]{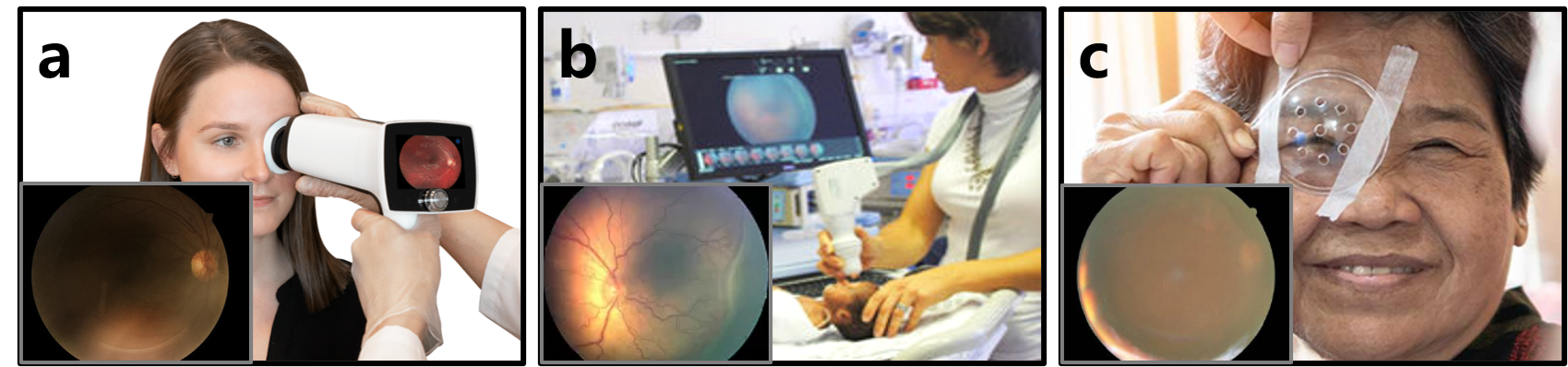}
        \par
    \end{centering}
\caption{Fundus photography in clinics. (a) Portable fundus photography, (b) fundus photography for infancy, (b) fundus photography for cataract patients.}
\vskip -5pt
\label{fig:application_datascarcity}
\end{figure}

For instance, portable fundus photography (Fig.~\ref{fig:application_datascarcity} (a)) is vulnerable to environmental interferences~\cite{palermo2022sensitivity}, while patient-related issues, such as infant fixation difficulties (Fig.~\ref{fig:application_datascarcity} (b)) and media opacity due to cataracts (Fig.~\ref{fig:application_datascarcity} (c)), also degrade fundus image quality~\cite{coyner2019automated}.
As a result, fundus photography allows clear critical structures in less than 49.6\% of patients \cite{chen2023quality}, and roughly 21\% of clinical fundus images are unsuitable for intelligent diagnostic systems \cite{beede2020human}.
This image degradation further introduces uncertainties in diagnostic decisions, underscoring the importance of guaranteeing high-quality fundus images for preventing vision impairment~\cite{liu2022understanding}.

\begin{figure*}[tbp]
    \begin{centering}
        \includegraphics[width=\linewidth]{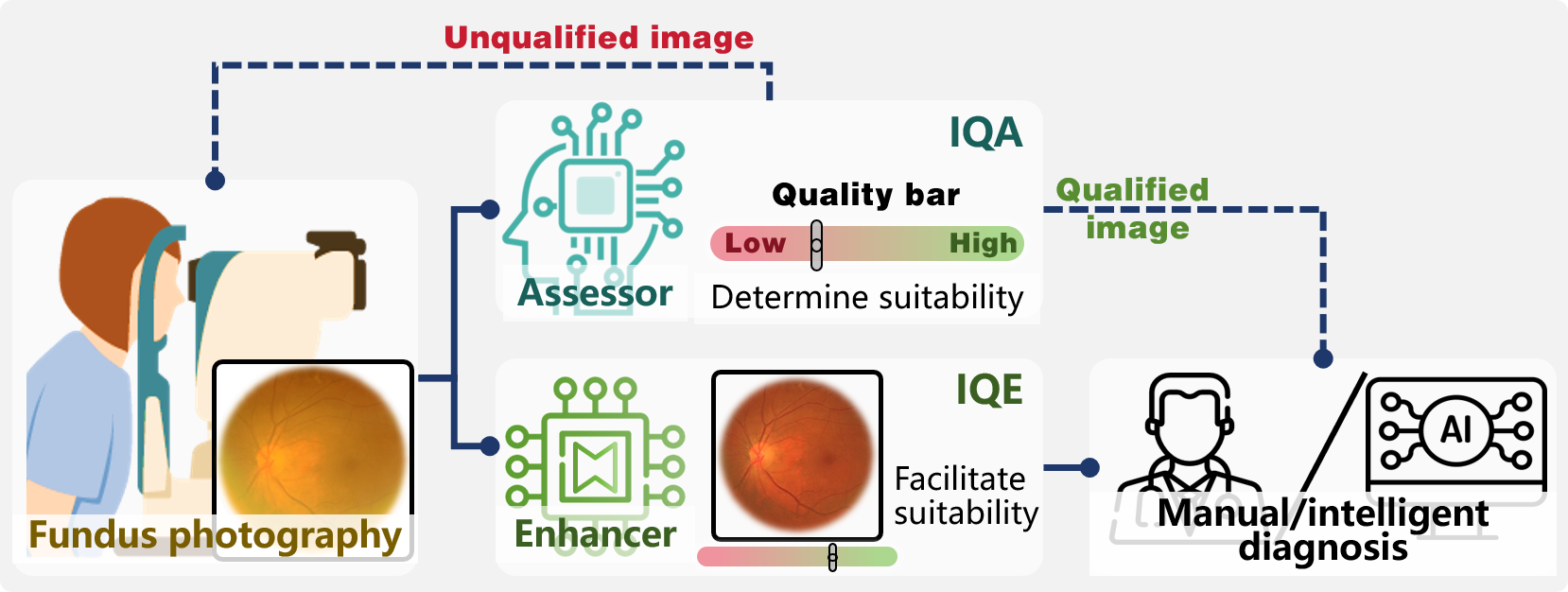}
        \par
    \end{centering}
\caption{The functions of IQA and IQE in guaranteeing high-quality fundus images.
IQA assesses the perceptual quality of fundus images to determine their suitability for diagnosing diseases.
A qualified image is directed towards diagnosis, while an unqualified one necessitates recapture.
IQE enhances the perceptual quality of fundus images to facilitate their suitability for diagnostic purposes.}
\vskip -5pt
\label{fig:defination}
\end{figure*}

In the pursuit of high-quality fundus images, IQA and IQE algorithms have been developed to aid in fundus image acquisition~\cite{das2023feasibility,raj2019fundus,zhang2024computational}.
Illustrated in Fig.~\ref{fig:defination}, IQA assists in the selection of fundus images by assessing their perceptual quality~\cite{raj2019fundus}, while IQE improves the perceptual quality of fundus images to facilitate their suitability for diagnostic purposes~\cite{zhang2024computational}.
Moreover, these techniques have been incorporated into intelligent diagnostic systems to boost the reliability of diagnostic conclusions~\cite{liu2023deepfundus}.
Despite the advancements in IQA and IQE, introducing these techniques into practical applications remains troublesome, with many well-designed algorithms encountering deployment challenges in clinical scenarios.
Therefore, there exists an urgent requirement for a comprehensive review of current studies and applications concerning fundus IQA and IQE to inform future research directions, especially with regards to clinical deployment.

\begin{figure}[tbp]
    \begin{centering}
        \includegraphics[width=1\linewidth]{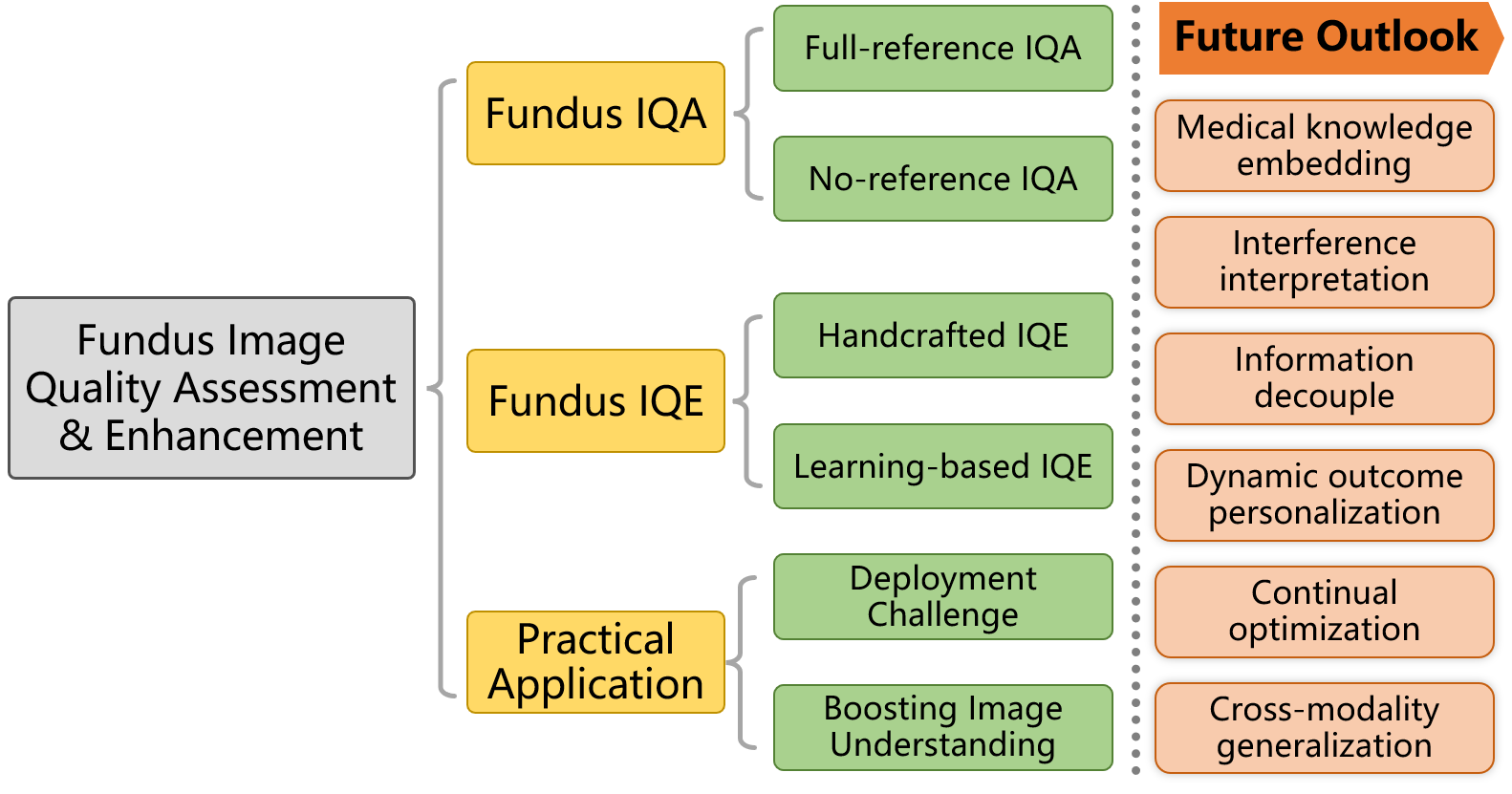}
        \par
    \end{centering}
    %\vspace{-2.5ex}
\caption{Manuscript taxonomy}
\vskip -5pt
\label{fig:taxonomy}
\end{figure}

While Li et al. \cite{li2021applications} and Iqbal et al. \cite{iqbal2022recent} have presented comprehensive reviews on fundus image analysis, their discussion of IQA and IQE is limited.
Regarding IQA, Raj. et al. \cite{raj2019fundus} offered a detailed survey, though it is now outdated.
Recently, Shi et al. \cite{shi2022assessment} and Gon{\c{c}}alves et al. \cite{gonccalves2024image} compared the performance of fundus IQA algorithms across datasets, and Volkov et al.\cite{volkov2024fundus} briefly summarized the strengths and weaknesses of fundus IQA algorithms.
In the realm of IQE, Bindhya et al. \cite{bindhya2020review} compared enhancement and denoising techniques in fundus images for diabetic retinopathy (DR), and Balashunmugam et al. \cite{balashunmugam2023image} provided a brief review of the technological paradigms in fundus IQE.
More recently, Zhang et al. \cite{zhang2024computational} reviewed fundus image restoration techniques, focusing on their specific mathematical models and performance.
Although these previous surveys provided individual reviews on IQA and IQE techniques, it is more logical to review their algorithms and applications cohesively due to the intimate relationship between IQA and IQE.

To address this void, this paper offers a comprehensive review that concentrates on the advancement of fundus IQA and IQE techniques from an integrated perspective. As shown in Fig.~\ref{fig:taxonomy}, our objective is to methodically explore the algorithmic paradigms and advancements in IQA and IQE alongside their application studies.
Specifically, we commence by delineating the fundamentals of the fundus photography imaging system and the associated interferences. Subsequently, we conduct a systemic review of paradigms and progress in IQA and IQE algorithms, where IQA algorithms are summarized according to their reliance on references, and IQE algorithms are categorized into handcrafted and learning-based paradigms. Following this, we examine the challenges and strategies for deploying IQA and IQE to boost fundus image understanding. Finally, we provide insights into potential future directions in this field.

\section{Fundus Imaging System and Interference}
Fundus photography, a non-invasive imaging method utilized to capture intricate images of the eye's posterior segment, is frequently employed in both standard eye examinations and specialized ophthalmic assessments.
The technique of fundus photography involves illuminating the eye's interior and recording the reflected light to generate a fundus image, akin to photographing any object but with specialized optics and illumination tailored to the distinctive requirements of imaging the eye's internal structures.

%TODO:图里加一些abc标注，结合标注分段描述成像过程
% As shown in Fig.\ref{fig:cfp_principle}, the principle of color fundus photography involves optical systems, photoelectric conversion, and digital image processing. 
% \subsection{Imaging System for Fundus Photography}

\begin{figure}[tbp]
    \begin{centering}
        \includegraphics[width=\linewidth]{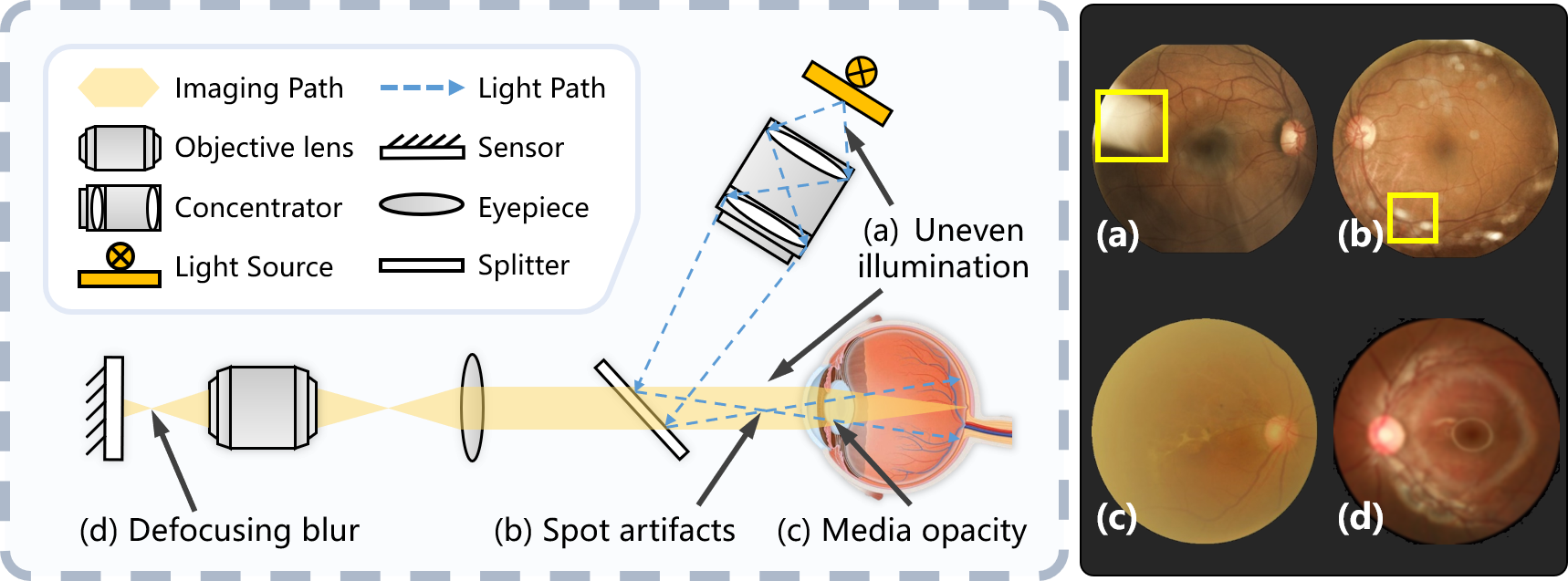}
        \par
    \end{centering}
    %\vspace{-2.5ex}
\caption{Imaging system and interference of fundus photography.}
\vskip -5pt
\label{fig:cfp_principle}
\end{figure}

Fig.~\ref{fig:cfp_principle} illustrates the fundamental principle of the imaging system used in fundus photography:

1) A gentle light source within the fundus camera projects light through the pupil, traversing the cornea and lens to illuminate the fundus. 
Notably, uneven illumination can occur during this procedure due to fluctuations in the light source or surrounding lighting conditions, leading to underexposure or overexposure in images, as depicted in Fig.~\ref{fig:cfp_principle} (a).

2) Variations in light reflection and absorption by retinal structures create the patterns of brightness and color captured by the fundus camera. Reflected light, traveling back through the eye's optics, is focused onto the camera's sensor to form the fundus image. 
Throughout this process, obstructions like eyelashes or dust can introduce artifacts, appearing as spot artifacts (shown in Fig.~\ref{fig:cfp_principle} (b)), while media opacities, such as cataracts, can scatter and attenuate light, resulting in blurry or low-contrast images, as illustrated in Fig.~\ref{fig:cfp_principle} (c).

3) The light reflected to create the fundus image on the camera's sensor is precisely focused by a series of lenses. If the patient fails to maintain proper eye fixation during imaging, a defocusing blur can emerge within the image, as demonstrated in Fig.~\ref{fig:cfp_principle} (d).

Due to the efficient imaging principle, fundus photography is embraced for the benefits of cost-effectiveness, user-friendly, and portability, rendering it a prevalent imaging technique for large-scale screening and clinical examinations. 
However, the interferences within the imaging system lead to quality degradation, a common issue encountered in clinical fundus image collections.
To maintain high-quality fundus images, the deployment of IQA and IQE algorithms is necessary.

% \begin{figure}[!t]
%     \begin{centering}
%         \includegraphics[width=0.9\linewidth]{images/DegradationEx.png}
%         \par
%     \end{centering}
%     %\vspace{-2.5ex}
% \caption{Typical degradation interference in fundus photography.}
% \vskip -5pt
% \label{fig:DegradationEx}
% \end{figure}

\section{Fundus IQA}
% 背景，偏医学
In clinical scenarios, IQA is vital for fundus examination and diagnosis to eliminate uncertainties stemming from unqualified images.
Regarding fundus images, image quality generally pertains to the image's suitability for downstream tasks, notably its effectiveness for diagnosis~\cite{jiang2022quality}. 
As a result, clinicians frequently conduct subjective IQA through manual labeling~\cite{raj2019fundus}. 
Nevertheless, subjective assessment is plagued by high costs and slow processing speeds~\cite{chen2023quality}, prompting a surge of interest in the development of objective IQA algorithms based on artificial intelligence.

In terms of evaluation standards, IQA algorithms can be categorized into two main types: \textbf{full-reference} and \textbf{no-reference}.
Full-reference IQA (FR-IQA) involves a direct comparison between the image being assessed and an ideal reference image.
No-reference IQA (NR-IQA), also referred to as Blind IQA (BIQA), diverges from FR-IQA in that NR-IQA methods do not necessitate a reference image as input.

\vspace{1cm}
\subsection{Full-reference IQA}
% 在reference时，用于比较一张图片与另一张确定为高质量图片的质量
Full-reference Image Quality Assessment (FR-IQA) calculates a quality score for a target image by comparing it to a given reference image.
The use of a reference image allows FR-IQA to provide specific, desired assessment metrics.
The numerical and learning-based FR-IQA methods are summarized.

\subsubsection{Numerical FR-IQA Metrics}
% 数值性指标：
%% MSE类
%% SSIM类
% 基于学习的指标：量太少，暂时不分 

Numerical FR-IQA metrics do not require model training and rely solely on static pixel comparisons, offering quantitative metrics for IQA. 
While these metrics are generally straightforward and widely applicable, in medical applications, they are primarily utilized to gauge the effects of image processing algorithms (like denoising, enhancement, reconstruction, etc.) on image quality due to their reliance on associated reference images.

\begin{figure}[tbp]
    \begin{centering}
        \includegraphics[width=0.9\linewidth]{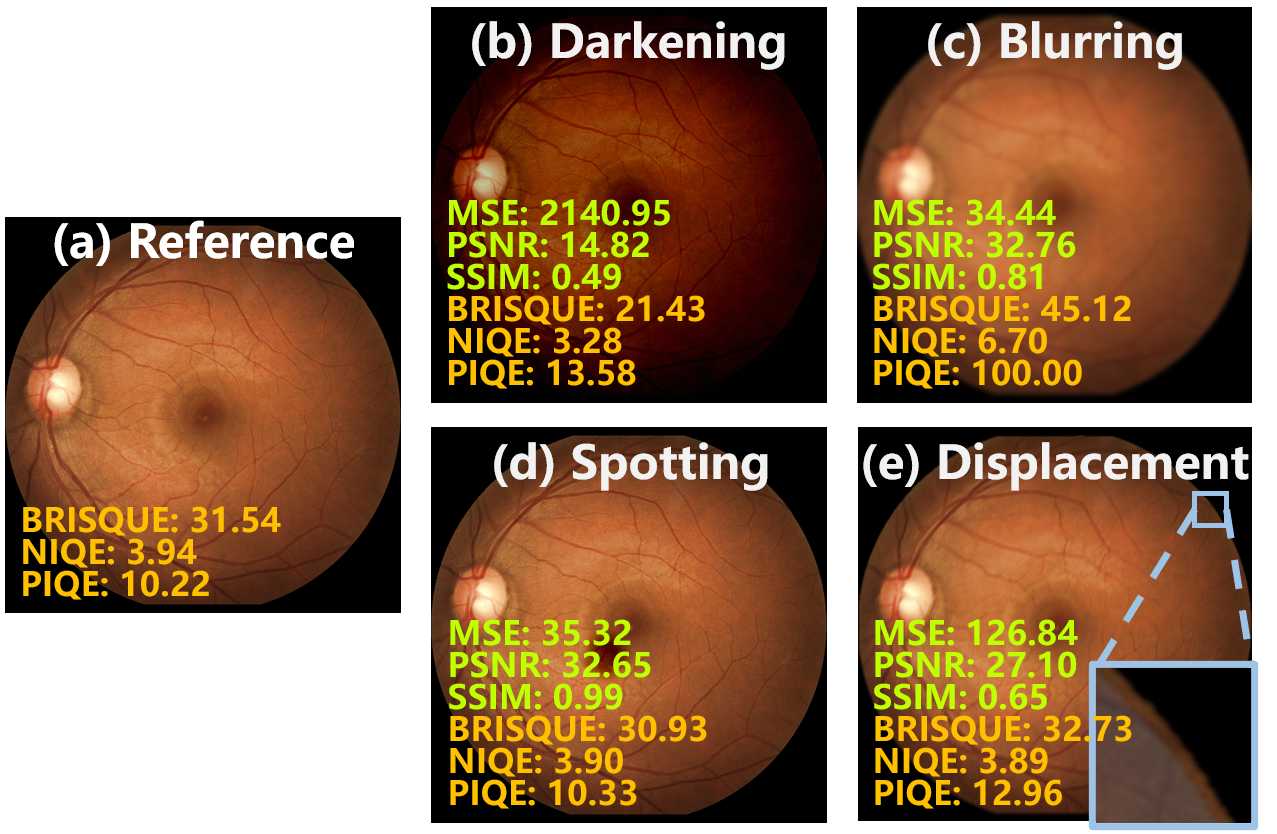}
        \par
    \end{centering}
    %\vspace{-2.5ex}
\caption{Exhibition of numerical \textcolor[rgb]{0.75,1,0}{FR-} and \textcolor[rgb]{1,0.75,0}{NR-IQA} metrics on images affected by (b) darkening, (c) blurring, (d) spotting, and (e) displacement compared to the reference (a).
Lower MSE scores, along with higher PSNR and SSIM scores indicate higher image quality.}
\vskip -5pt
\label{fig:numerical}
\end{figure}

A comparison of numeric FR-IQA metrics is presented in Fig.~\ref{fig:numerical}. 
These metrics can be broadly categorized into two types:  information fidelity of images at the pixel-level and in the structure.

\noindent \textbf{Pixel-level Fidelity}

Mean Square Error (MSE) and Peak Signal-to-Noise Ratio (PSNR) are the most common information fidelity assessments at the pixel-level.

Given a target image $I$ and a reference image $R$, with the dimensions $m \times n$. 
MSE is defined as:
% The most direct method of full-reference comparison involves assessing signal fidelity, with mean square error (MSE) and peak signal-to-noise ratio (PSNR) being the commonly utilized metrics~\cite{sara2019image}. Given two images, $I$ and $K$, each of dimensions $m \times n$, where $I$ serves as the reference image and $K$ is the image whose quality is to be assessed. Then, MSE is defined as:

\begin{equation}
MSE(I,R)=\dfrac{1}{mn}\sum_{x=0}^{m-1}\sum_{y=0}^{n-1}[I(x,y)-R(x,y)]^2.
\end{equation}

Based on MSE, PSNR is defined as:

\begin{equation}
PSNR(I,R)=10\cdot \lg(\dfrac{MAX_I^2}{MSE}),
\end{equation}
where $MAX_I$ denotes the maximum value of the pixels in $I$.

Lower MSE scores and higher PSNR scores indicate a closer similarity to the reference, reflecting higher image quality.
As exhibited in Fig.~\ref{fig:numerical}, MSE is overly sensitive to certain interference. 
For instance, the darkening interference in Fig.~\ref{fig:numerical} (b) results in an MSE score exceeding 2140.
This instability makes MSE unsuitable for objective and consistent IQA, particularly for fundus images.
In contrast, PSNR offers greater stability and is therefore more frequently employed for fundus IQA in related works \cite{li2022annotation,Agenericfundusimageenhancementnetwork}.
Because quantifying similarity at the pixel-level, both MSE and PSNR exhibit sensitivity to darkening and displacement (Fig.~\ref{fig:numerical} (b) and (e)), while insensitivity to blurring and spotting (Fig.~\ref{fig:numerical} (c) and (d)).

\noindent \textbf{Structure Fidelity}

Structural Similarity Index Measure (SSIM)~\cite{wang2004image} is a representative FR-IQA metric of information fidelity concerning image structure. SSIM considers aspects including luminance, contrast, and structure, rather than solely focusing on pixel-level distinctions. 

SSIM is defined as:

\begin{equation}
SSIM(I,R)=[l(I,R)]^\alpha\cdot [c(I,R)]^\beta\cdot[s(I,R)]^\gamma,
\end{equation}
where $l(I,R)$, $c(I,R)$ and $s(I,R)$ denote the similarity in luminance, contrast, and structure, respectively, between corresponding regions of interest (ROI) in $I$ and $R$.

% \begin{equation}
% l(I,R)=\dfrac{2\mu_I\mu_R+C_1}{\mu_I^2+\mu_R^2+C_1},
% \end{equation}

% \begin{equation}
% c(I,R)=\dfrac{2\sigma_I\sigma_R+C_2}{\sigma_I^2+\sigma_R^2+C_2},
% \end{equation}

% \begin{equation}
% s(I,R)=\dfrac{\sigma_{IR}+C_3}{\sigma_I\sigma_R+C_3},
% \end{equation}
% where $\mu$ and $\sigma$ denote the mean value and standard deviation of pixels in the ROI, and $\sigma_{IR}$ means the covariance. 
% $C_1$, $C_2$, and $C_3$ are constants used to avoid division by zero in the denominator.

A higher SSIM score refers to higher image quality.
Compared to MSE and PSNR, SSIM captures global structural variances more effectively and aligns more closely with the judgment of the human visual system (HVS).
Therefore, SSIM reflects the quality degradation from blurring (Fig.~\ref{fig:numerical} (c))
Nonetheless, according to Fig.~\ref{fig:numerical} (d), SSIM may not exhibit sufficient sensitivity to local alterations, which could be crucial medical details.
Similar to MSE and PSNR, SSIM, being an FR-IQA metric, is sensitive to the displacement between the target image and the reference (Fig.~\ref{fig:numerical} (e)), even when the displacement has little impact on image quality.

Furthermore, SSIM has various derived IQA metrics aimed at improving its performance from different perspectives. To address the resolution diversity in IQA, the multi-scale SSIM (MS-SSIM) conducts an iterative analysis across multiple scales, enabling the IQA across different resolutions~\cite{wang2003multiscale}. Additionally, leveraging the framework of MS-SSIM, the information-weighted SSIM (IW-SSIM) was introduced to improve assessment accuracy by applying a weighted combination of local SSIM indices based on the information distribution throughout the entire image~\cite{wang2010information}. Moreover, Feature Similarity (FSIM) takes into account the HVS's focus on low-level distinctive features by comparing phase congruency and gradient magnitude between images~\cite{zhang2011fsim}.

\subsubsection{Learning-based FR-IQA Methods}
The evolution of deep learning has spurred the development of learning-based FR-IQA methods.
Leveraging the strengths of deep learning, these methods excel at extracting relevant features from specific image categories, facilitating tailored comparisons between reference and target images to satisfy distinct medical criteria. Consequently, they exhibit enhanced performance compared to traditional methods in achieving quality evaluations that more closely align with human subjective judgments, particularly within specialized task contexts or datasets.

In contrast to numerical FR-IQA metrics, learning-based FR-IQA methods typically utilize the reference image as an auxiliary conditional input, rather than conducting a pixel-level comparison between the reference and target images. 
A backbone network designed to handle dual inputs is often utilized, where both the reference and target images are simultaneously loaded to approximate an IQA function based on quality labels.
Bosse et al. \cite{bosse2017deep} introduced a CNN-based FR-IQA model with a multi-input architecture to merge features extracted from both the reference and target images, incorporating weighted averages to emphasize local regions.
Fang et al. \cite{fang2020perceptual} introduced an FR-IQA method that integrates CNN with transformer architecture to enhance image feature perceptual capabilities and evidenceable performance improvement.

\subsection{No-reference IQA}
% 数值性指标
% 端到端的指标
% 基于综合评估的指标
%% 基于全局特征设计的综合评估
%% 基于局部或像素级特征涉及的综合评估

% NRIQA或称BIQA，reference时直接评估单张图质量，得到一个数字（连续或离散）
In contrast to FR-IQA, NR-IQA directly provides a quality score or rating for an input image without requiring an ideal reference image. This independence from reference images enables NR-IQA methods to be utilized in a broader range of situations. In medical scenarios, NR-IQA can act as a standard for the usability of images captured by medical devices, as well as deliver customized evaluations for IQE modules integrated within fundus photography platforms.

NR-IQA methods are summarized in three categories: 1) numerical NR-IQA metrics, 2) end-to-end NR-IQA methods, and 3) comprehensive NR-IQA methods.

\subsubsection{Numerical NR-IQA metrics}
Numerical NR-IQA metrics conduct calculations based on specific characteristics or statistical information of images to quantify image quality. 
Compared to FR-IQA metrics, NR-IQA metrics are independent of reference images, offering benefits in terms of objectivity and repeatability.

% \begin{figure}[htbp]
%     \begin{centering}
%         \includegraphics[width=0.9\linewidth]{images/NRIQA_numerical_metrics.png}
%         \par
%     \end{centering}
%     %\vspace{-2.5ex}
% \caption{Exhibition of numerical NR-IQA metrics on fundus images.
% Lower scores indicate higher image quality.}
% \vskip -5pt
% \label{fig:nriqa_numerical}
% \end{figure}

The most commonly utilized numeric NR-IQA metrics are outlined below, with a performance comparison provided in Fig.~\ref{fig:numerical}.
Additionally, a list and value attributes of numerical FR- and NR-IQA metrics are presented in Table~\ref{tab:metrics}.

The representative NR-IQA metrics comprise:
\begin{itemize}
    \item \textbf{BRISQUE} (Blur/Repeat/Independent Que) slices the image into small chunks, learns the differences between typical and distorted images by analyzing local statistical features (e.g., gradients and contrast), and predicts new image quality based on these learned differences.
    \item \textbf{NIQE} (Natural Image Quality Evaluator) evaluates natural image quality by analyzing local and global statistical features (e.g., gradient, contrast) and applying machine learning methods to reflect image naturalness.
    \item \textbf{PIQE} (Perceptual Image Quality Evaluator) uses perceptual features, structure, and color distribution to simulate human vision for an accurate image quality assessment.
    % \item \textbf{FID} (Fréchet Inception Distance) uses the Inception network to extract image features, then calculates the mean and covariance of generated and natural image features, evaluating their difference using the Fréchet distance.
\end{itemize}

% \begin{table}[htbp]
% \centering
% \small
% \caption{List and value attributes of numerical FR- and NR-IQA metrics. 
% $\uparrow$ signifies higher values denote better quality, while $\downarrow$ indicates the opposite.}
% \label{tab:metrics}
% % \renewcommand{\arraystretch}{1.15}
% \begin{tabular}{|p{2.2cm}|p{5.8cm}|}
% \hline
% \textbf{Metric type} & \textbf{Numerical metric list and attributes} \\
% \hline
% \rowcolor[gray]{.90} FR-IQA
%  & MSE$\downarrow$, PSNR$\uparrow$, SSIM$\uparrow$ \\
%  \hline
% \rowcolor[gray]{.90} NR-IQA
%  & BRISQUE$\downarrow$, NIQE$\downarrow$, PIQE$\downarrow$ \\
%  \hline
% \end{tabular}
% \end{table}

\begin{table}[tbp]
\centering
\footnotesize
\caption{List and attributes of numerical FR- and NR-IQA metrics. 
$\uparrow$ signifies higher values denote better quality, while $\downarrow$ indicates the opposite.}
\label{tab:metrics}
\begin{tabular}{l|l|l|l}
\hline
Metric & Type & Trend & Aspect \\
\hline
\rowcolor[gray]{.90} MSE
 & FR & $\downarrow$ & Pixel-level fidelity \\
\hline
\rowcolor[gray]{.90} PSNR
 & FR & $\uparrow$ & Pixel-level fidelity \\
\hline 
\rowcolor[gray]{.90} SSIM
 & FR & $\uparrow$ & Pixel-level and structure fidelity \\
\hline 
\rowcolor[gray]{.90} BRISQUE
 & NR & $\downarrow$ & Gradients, contrast \\
\hline 
\rowcolor[gray]{.90} NIQE
 & NR & $\downarrow$ & Gradients, contrast, naturalness\\
\hline 
\rowcolor[gray]{.90} PIQE
 & NR & $\downarrow$ & Structure, noise, color\\
\hline 
% \rowcolor[gray]{.90} FR-IQA
%  & MSE$\downarrow$, PSNR$\uparrow$, SSIM$\uparrow$ \\
%  \hline
% \rowcolor[gray]{.90} NR-IQA
%  & BRISQUE$\downarrow$, NIQE$\downarrow$, PIQE$\downarrow$ \\
%  \hline
\end{tabular}
\end{table}

As depicted in Fig.~\ref{fig:numerical}, NR-IQA metrics allow us to quantify quality scores for individual images directly.
Compared to the original image in Fig.~\ref{fig:numerical} (a), NR-IQA metrics may not effectively capture the darkening degradation in Fig.~\ref{fig:numerical} (b). This limitation arises from their initial design for natural images, making them less suitable for identifying illumination variations in fundus images.
In contrast to FR-IQA metrics, NR-IQA metrics exhibit greater sensitivity to blurring degradation in Fig.~\ref{fig:numerical} (c), while robust to displacement in Fig.~\ref{fig:numerical} (e) due to their independence from reference images.
Due to the limited impact of global image characteristics, NR-IQA metrics have constraints in effectively reflecting spotting degradation in Fig.~\ref{fig:numerical} (d).

\subsubsection{End-to-end NR-IQA methods}
\begin{figure}[htbp]
    \begin{centering}
        \includegraphics[width=\linewidth]{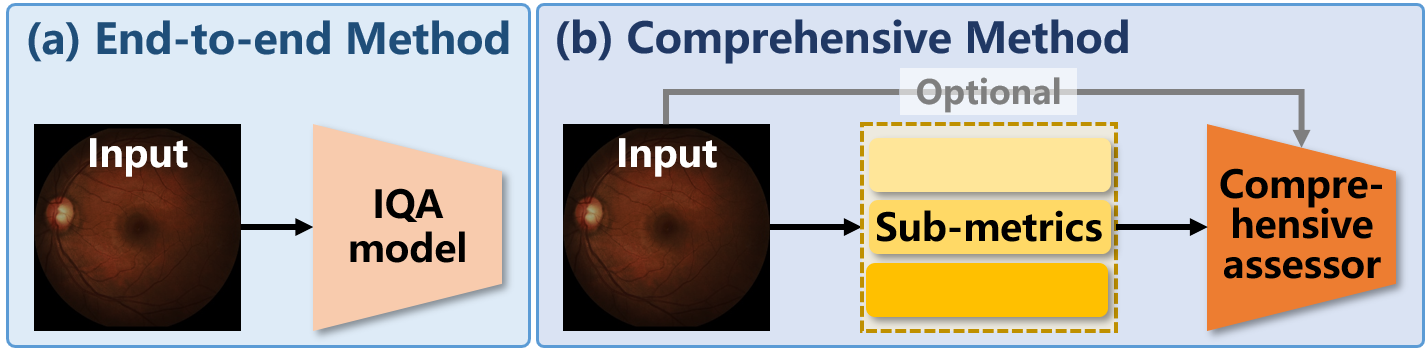}
        \par
    \end{centering}
    %\vspace{-2.5ex}
\caption{Scematics of naive end-to-end methods and comprehensive evaluation methods.}
\vskip -5pt
\label{fig:nriqa_scematics}
\end{figure}

% 端到端，直接输出分数，对中间结果没有进行可解释性的（明确与某些质量指标相关）约束
As depicted in Fig.~\ref{fig:nriqa_scematics} (a), end-to-end NR-IQA methods generate evaluation ratings directly using a learning-based classification or regression model.

% Naive end-to-end methods in IQA tasks refer to approaches where the model directly outputs evaluation scores without imposing explicit interpretability constraints on intermediate results. As shown in Fig.\ref{fig:nriqa_scematics}.(a), these methods typically use an image classifier or regression model as the backbone and do not involve specially designed architectures for specific tasks.

Denote the model in end-to-end NR-IQA methods as $F_\theta$, where $\theta$ is learnable parameters. Given training data $\{I,L\}$, where $I$ represents target images and $L$ denotes quality labels (either discrete for classification or continuous for regression), the optimization objective of $F_\theta$ can be formulated as:
\begin{equation}
\underset{\theta}{arg\,min}\mathcal{L}_{IQA}[F_\theta (I),L],
\end{equation}
where $\mathcal{L}_{IQA}[\cdot, \cdot]$ denotes the loss function for classification (such as cross-entropy loss) or regression (such as MSE loss).

\noindent\textbf{Classification-oriented models}

In prior research, classification-oriented convolutional neural networks (CNNs) have commonly been employed as IQA models. 
Straightforward approaches involve implementing typical CNNs to classify quality labels for fundus images, and more advanced approaches boost CNN performance in IQA by further introducing additional modules.
Specifically, Zago et al. \cite{zago2018retinal} implemented a fundus IQA model by directly utilizing the Inception V3 model fine-tuned on NR-IQA datasets.
To address the challenge of small sample sizes and potential inconsistencies between training and testing distortions in IQA tasks, Zhu et al. \cite{zhu2020metaiqa} proposed a meta-learning-based NR-IQA method, enhancing model generalization against unknown distortions.
Liu et al. \cite{liu2021small} proposed an NR-IQA model leveraging resampling and gcForest methods to tackle issues of sample imbalance and limited data in IQA tasks.
By disentangling image quality degradation factors, Shi et al. \cite{shi2022assessment} proposed a guideline for differentiating damage and artifacts in fundus images to enhance the performance of the NR-IQA model.
Lin et al. \cite{lin2023domain} introduced an adaptive IQA framework using a teacher-student model to conduct knowledge distillation across IQA datasets. 

\noindent\textbf{Regression-oriented models}

To offer quantitative IQA outcomes, regression-oriented CNNs have been suggested.
Drawing from multivariate linear regression fused with various CNN features, a quantitative NR-IQA model was presented for fundus images (Raj et al., 2020).
Zhang et al. \cite{zhang2021uncertainty} introduced uncertainty into the NR-IQA model to align assessment scores with varying standards across different databases, incorporating uncertainty regularization to mimic human decision processes.

\begin{table*}[htbp]
\centering
\footnotesize
\caption{Representative fundus IQA models.}
\label{tab:iqa}
\begin{tabular}{l|l|l|l}
\hline
Article & Quality Perception & Contribution & Dataset \\
\hline

\rowcolor[gray]{0.9} \cite{zago2018retinal} & End-to-end classification & Fine-tune CNN for fundus IQA & DRIMDB \\
\hline
\rowcolor[gray]{0.9} \cite{shi2022assessment} & End-to-end classification & Random forest with CNN features & / \\
\hline
\rowcolor[gray]{0.9} \cite{lin2023domain} & End-to-end classification & Domain adaptation by knowledge distillation & DRIMDB \\
\hline

\rowcolor[gray]{0.9} \cite{wang2015human} & Illumination, color, focus, contrast & HVS prior-based features & DRIMDB \\
\hline
\rowcolor[gray]{0.9} \cite{costa2017eyequal} & Classification with quality map & Explainable patch-level IQA & DRIMDB \\
\hline
\rowcolor[gray]{0.9} \cite{chalakkal2019quality} & Illumination, sharpness, homogeneity, vision field & Statistic and segmentation features & DRIMDB \\
\hline
\rowcolor[gray]{0.9} \cite{fu2019evaluation} & Multiple color spaces & EyeQ dataset color-space features & EyeQ \\
\hline
\rowcolor[gray]{0.9} \cite{kanimozhi2020fundus} & Illumination, naturalness, structure & Adaptive optimization & DRIMDB \\
\hline
\rowcolor[gray]{0.9} \cite{biswas2020grading} & Illumination, intensity, structure & Multichannel segmentation features & / \\
\hline
\rowcolor[gray]{0.9} \cite{shen2020domain} & Artifact, clarity, vision field, structure & Explainable quality map \& domain adaptation & Private \\
\hline
\rowcolor[gray]{0.9} \cite{dai2021deep} & Artifact, clarity, vision field & IQA coupled with diagnosis and structure & Private \\
\hline
\rowcolor[gray]{0.9} \cite{xu2022dark} & Dark and bright channel prior information & Prior-guided assessment & EyeQ \\
\hline
\rowcolor[gray]{0.9} \cite{liu2023deepfundus} & Illumination, clarity, position & Sub-IQA models & Private \\
\hline
\rowcolor[gray]{0.9} \cite{xu2023deep} & Salient structure & Large- and small-size segmentation features & EyeQ \\
\hline
\rowcolor[gray]{0.9} \cite{yi2023label} & Prior, frequency, CLIP features & Foundation model features & EyeQ \\
\hline
\end{tabular}
\end{table*}

\subsubsection{Comprehensive NR-IQA Methods}
As shown in Fig.\ref{fig:nriqa_scematics} (b), comprehensive NR-IQA methods are also based on learning models, but introduce intermediate sub-metrics into the quality assessment process.
The comprehension of these sub-metrics is considered to improve the generalizability and reliability of NR-IQA algorithms. 

% The other type of NR-IQA method is based on comprehensive evaluation. As shown in Fig.\ref{fig:nriqa_scematics}.(b), comprehensive evaluation introduces additional sub-metrics into the intermediate quality assessment process, which is assumed to improve the accuracy and generalizability of NR-IQA algorithms.

The comprehensive NR-IQA method incorporates sub-metric models $E^{n}_{\phi}$ alongside the assessor $F'_{\theta}$.
The training data is represented as $\{I,L,[d_n]\}$, where $[d_n]$ signifies the labels of the sub-metrics.
% Generally, comprehensive evaluation methods are more complex than end-to-end methods. This approach incorporates a sub-metrics evaluation model $E_{\theta_E}$ alongside the IQA model $F_{\theta_F}$. The training data is $\{x,y,(d_0,d_1, \cdots, d_{k-1})\}$, where $(d_0,d_1, \cdots, d_{k-1})$ are sub-metrics labels. 
The optimization procedure involves training both $E^{n}{\phi}$ and $F'{\theta}$. The objective for the sub-metric models is given by:
%TODO:这部分方法需要加个示意图，展示一下现在算法怎么结合具体指标的
\begin{equation}
\underset{\phi}{arg\,min} \mathcal{L}_{sub}[E^n_{\phi}(I),[d_n]], 
\end{equation}
where $\mathcal{L}_{sub}[\cdot, \cdot]$ is the loss function used to align prediction and label of sub-metrics.

Subsequently, the objective for the comprehensive assessor is defined as:
\begin{equation}
\underset{\theta}{arg\,min} \mathcal{L}_{IQA}[F'_{\theta}(E^n_{\phi}(I),L],
\end{equation}

The comprehensive NR-IQA methods usually concentrate on two aspects of sub-metrics: 

1) Visual effects are the core of IQA, whose metrics aim to quantify how distortions affect the perceived visual quality of an image, such as sharpness, contrast, noise, color accuracy, and artifacts.

2) Semantic content pertains to the perceptibility and completeness of medical contents and task-related clues.

% Several prevalent sub-metrics related to fundus IQA are summarized in Table \ref{tab:sub_metrics}.

\noindent\textbf{Visual Effects}

Visual effects, in the context of IQA, are quantifiable characteristics of an image that contribute to its perceived visual quality. 
These characteristics are frequently employed as sub-metrics to depict distortions introduced during image processing, transmission, or acquisition.

Handcrafted sub-metrics are convenient to obtain for comprehensive NR-IQA.
Moorthy et al. \cite{moorthy2011blind} devised a two-stage NR-IQA approach that amalgamated various statistical features of different degradations to produce a comprehensive evaluation.
Kanimozhi et al. \cite{kanimozhi2020fundus} derived specific quality scores for image illumination and naturalness to establish an interpretable fundus IQA model.
Xu et al. \cite{xu2022dark} employed dark channel and bright channel maps to facilitate NR-IQA based on the imaging characteristics of fundus photography.

With the advancement of deep learning, learning-based sub-metrics are rapidly developed.
Costa et al. \cite{costa2017eyequal} proposed a patch-based IQA strategy for fundus photography, where each image patch is independently assessed for quality, and an overall score is computed through weighted average pooling.
Pan et al. \cite{pan2018blind} recommended aligning the intermediate features from various FR-IQA models to create an NR-IQA model.
Drawing on a multi-color spatial fusion network, Fu et al. \cite{fu2019evaluation} proposed a comprehensive fusion of multiple visual attributes for fundus NR-IQA.
Diverse image quality attributes and category labels in smartphone images have been leveraged to boost NR-IQA~\cite{fang2020perceptual}.
Dai et al. \cite{dai2021deep} developed a multitasking fundus color photo processing algorithm for DR, facilitating simultaneous image quality assessment, retinopathy classification, and DR grading, thus catering to diverse clinical requirements. 
Ou et al. \cite{ou2021sdd} utilized similarity distribution distance as a pseudo-label for IQA tasks, improving the model performance.
Ma et al. \cite{ma2021blind} introduced an NR-IQA model mimicking HVS, leveraging GAN to generate similarity maps with the original image and utilizing multi-input networks.

\noindent\textbf{Semantic Content}

As semantic content is pivotal in the examination and diagnosis of medical images,  scholars have suggested that integrating semantic attributes into IQA would be judicious.
For fundus images, these semantic attributes are usually measured by the prior knowledge of fundus photography and downstream tasks related to anatomical structures, like segmentation.

Using prior knowledge of fundus photography, semantic content can be imposed in the NR-IQA models.
Li et al. \cite{li2018has} proposed an NR-IQA method based on semantic feature aggregation to mitigate perturbation effects during inference.
Kanimozhi et al. \cite{kanimozhi2020fundus} integrated sub-metrics of structural level with illumination and naturalness to improve the interpretability of IQA models.
Leveraging CLIP~\cite{radford2021learning}, Yi et al. \cite{yi2023label} proposed a semantics-aware contrastive learning model for NR-IQA in medical images.
Liu et al. \cite{liu2023deepfundus} introduced a comprehensive fundus IQA method that individually evaluates the position, illumination, and clarity of various fundus organs.

Segmentation modules are frequently employed to incorporate anatomical structures into NR-IQA methods.
Wang et al. \cite{wang2015human} integrated vessel segmentation outcomes into an NR-IQA model designed for fundus images.
In the NR-IQA approach proposed by~\cite{chalakkal2019quality}, a segmentation-based shot location evaluation was utilized to provide medically relevant assessments.
Biaswas et al. \cite{biswas2020grading} merged quality metrics and structural segmentation in NF-IQA models by incorporating luminance histogram features with vessel and optic disc segmentation results.
Shen et al. \cite{shen2020domain} employed local feature extraction and identification of key anatomical structures to consider fundus image artifacts, clarity, field definition, and segmentable factors in IQA.
An NR-IQA method~\cite{xu2023deep} that focuses on significant anatomical structures and lesions was developed using segmentation results of large-size structures like the optic disc and exudate, as well as small-size structures like blood vessels.

\subsection{Challenges and Insight}
Both FR-IQA and NR-IQA play pivotal roles in clinical data collection and IQE algorithm evaluation.
FR-IQA proves powerful when low- and high-quality fundus image pairs are available, such as the evaluation of enhancement, super-resolution, and compression algorithms. However, due to the challenges associated with collecting paired data, NR-IQA is more practical and commonly applied in clinical settings, particularly in aiding fundus photography.
Table~\ref{tab:iqa} summarizes the representative NR-IQA models.

Learning-based NR-IQA shows potential by automatically establishing assessment criteria and even sub-metrics based on provided training data.
However, challenges persist in developing NR-IQA methods for fundus images. Firstly, inconsistencies in quality annotation standards, such as variations in imaging conditions and subjective annotations, lead to divergent classification boundaries across fundus datasets. Secondly, many algorithms overlook the unique characteristics of fundus images, including vessel patterns, lesions, and macular structures, which diminishes their robustness and reliability.

Future research could explore several paths. Firstly, the development of adaptive IQA algorithms that can adapt to specific annotation standards by adjusting hyperparameters or utilizing prompts tailored to different datasets. Secondly, the creation of algorithms incorporating fundus-specific metrics to enhance sensitivity to crucial features like retinal lesions and blood vessels, potentially boosting diagnostic accuracy.

\section{Fundus IQE}
IQA aids in detecting quality degradation in clinical images and directing the re-acquisition necessary to ensure that image data meets clinical requirements. However, the repeated collection not only escalates workload and expenses but also poses challenges for certain patient demographics (such as infants and cataract patients) due to physical constraints in obtaining high-quality images through repeated sessions. Therefore, the introduction of IQE to restore image quality holds significance in improving the efficiency of clinical image acquisition, as well as advancing disease diagnosis and treatment~\cite{liu2023deepfundus}.
% \begin{figure}[tbp]
%     \begin{centering}
%         \includegraphics[width= \linewidth]{images/enhance/4-2-1-5.png}
%         \par
%     \end{centering}
%     %\vspace{-2.5ex}
% \caption{Collect high-quality fundus images from low-quality ones by IQE.}
% \vskip -5pt
% \label{fig:enhance-1}
% \end{figure}

% As depicted in Fig.~\ref{fig:enhance-1}, 
IQE can notably enhance the low-quality data, consequently diminishing the necessity for image re-acquisition. 
Hence, IQE facilitates the fundus image acquisition outcomes for particular patient cohorts, streamlining the acquisition procedure and cutting down associated expenses.

% As shown in Fig.~\ref{fig:enhance-1}, combined with the quality assessment results, image enhancement processing on the data that meets the "usable" standard can significantly improve the image quality, thus reducing the need for re-acquisition of images, improving the image acquisition effect for specific patient groups, simplifying the acquisition process and reducing related costs.

Following the development trajectory, fundus IQE can be broadly categorized into two phases: \textbf{handcrafted methods} and \textbf{learning-based methods}.
Moreover, within handcrafted methods, the paradigms can be segmented into pixel-based processing, image filtering, and statistical prior techniques.
On the other hand, learning-based methods can be decomposed into paired data-oriented models and unpaired-oriented models.

\subsection{Handcrafted IQE Methods}
% Hand-crafted methods for image enhancement can be categorized into three main groups based on their processing approaches: 1)Pixel-Based Image: Histogram Equalization, gray Level Transformation; 2) Filtering: Gaussian Filter, mean filtering, median filtering, guided filtering; 3) Statistical Prior: Green Channel, dark Channel. These techniques have been widely applied to enhance retinal fundus images, and the details are explained below.

    \subsubsection{Pixel-based Image Processing}
    Pixel-based image processing commonly employs histogram equalization and grayscale transformations to enhance image quality by recovering lost contrast through the redistribution of brightness values.

\noindent\textbf{Histogram Equalization}

Histogram equalization~\cite{HE} enhances the visibility of image details in overexposed or underexposed areas by remapping each pixel value according to the image's histogram. 
This histogram represents the distribution of pixel values in an image, illustrating brightness and contrast patterns.
By transforming the histogram to a more uniform distribution, intensity values are spread across a wider range, increasing the dynamic range.
This results in a more balanced and visually appealing image, especially when the original image has a narrow range of intensities or suffers from poor contrast.

Given an image $I(x,y)$ comprising $N$ pixels with $L$ intensity levels, let the histogram value for intensity level $l$ be represented as $H(l)$. The cumulative distribution function (CDF) in histogram equalization is defined as $CFD(l)= \sum_{l=0}^{L-1}H(l)/N$. The equalized image $I'(x,y)$ is thus given by
\begin{equation}
    I'(x,y)= round[(L-1)\cdot \sum_{l=0}^{I(x,y)}H(l)/N].
    \label{HE} 
\end{equation}

\begin{figure}[tbp]
    \begin{centering}
        \includegraphics[width=\linewidth]{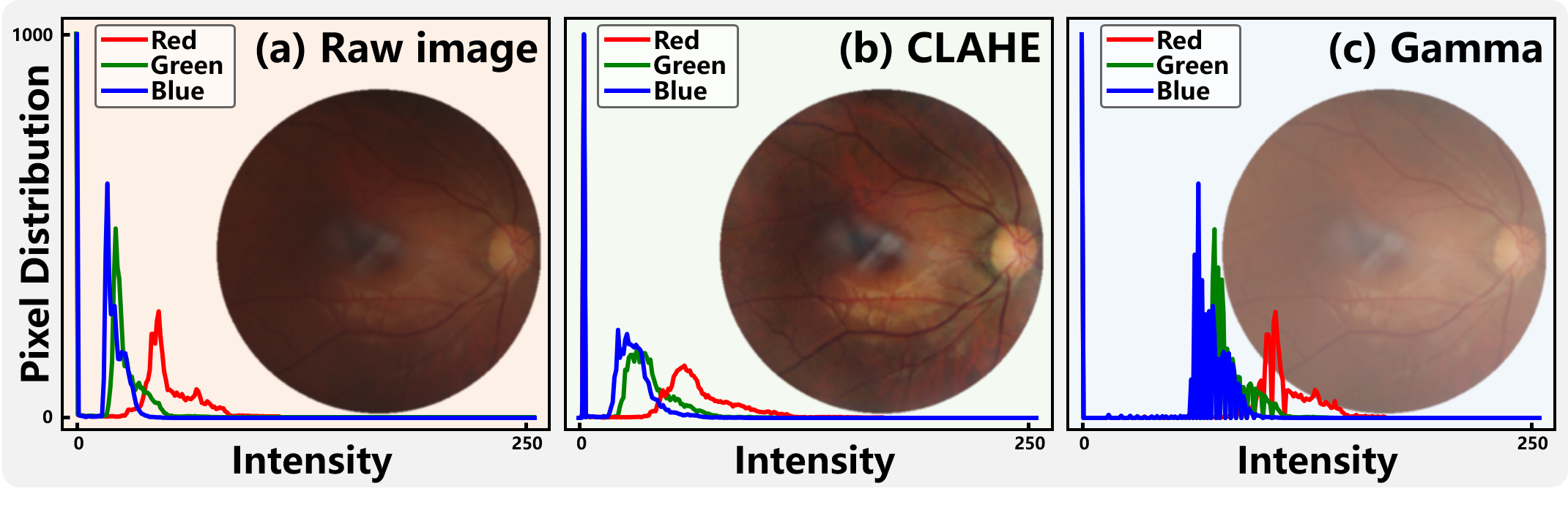}
        \par
    \end{centering}
    %\vspace{-2.5ex}
\caption{Example histograms and images normalized by CLAHE and Gamma correction. (a) the raw image, (b) the image normalized by CLAHE, and (c) the image normalized by Gamma correction.}
\vskip -5pt
\label{fig:enhance-2}
\end{figure}

However, the native histogram equalization may reduce image details. Subsequently, several improvements have been proposed. Adaptive Histogram Equalization (AHE) was introduced to achieve local contrast enhancement, but it may amplify noise excessively. Contrast-Limited Adaptive Histogram Equalization (CLAHE)~\cite{reza2004realization} enhances local contrast by dividing the image into small regions called tiles, as shown in Fig.~\ref{fig:enhance-2} (b). 
In CLAHE, histogram equalization is implemented with a clipping threshold to curb noise escalation within each tile, which is defined as
\begin{equation}
H^{\prime}(l)= \begin{cases}H(l), & \text { if } H(l) \leq C_{\text {clip }} \\ 
C_{\text {clip }}, & \text { if } H(l)>C_{\text {clip }}\end{cases}
\end{equation}
where $C_{\text {clip }}$ is the maximum allowable count for each intensity value.
CLAHE is implemented by apply $H^{\prime}(l)$ into Eq.~\ref{HE}.

Histogram equalization has been extensively employed in Fundus IQE.
Gandhama et al. \cite{gandhamal2017local} leveraged histogram equalization to generate well-defined and sharp edges between adjacent tissues by enhancing the difference between the minimum and maximum grayscale values.
Through comparing various histogram equalization derivatives, Salem et al. \cite{salem2019medical} validated the superior enhancement capabilities of CLAHE in fundus images.
To circumvent challenges related to sub-image selection and clipping parameters, AC-CLAHE and FA-CLAHE have been proposed in~\cite{RetinalfundusimageenhancementusingadaptiveCLAHE} to adaptively select the number of sub-images and clipping thresholds.
In \cite{kim2024fundus}, CLAHE was utilized as a preprocessing step to gather high-quality references from clinical fundus images for the training of IQE models.

\noindent\textbf{Grayscale transformation} 

Grayscale transformation involves adjusting the grayscale value of each pixel within an image to enhance its visual quality or extract specific features. 
In recent years, gamma correction~\cite{farid2001blind} has been frequently employed to impose the efficacy of grayscale transformations, as exhibited in Fig.~\ref{fig:enhance-2} (c).

Nonetheless, grayscale transformation can impact image brightness and lead to the loss of edge details. 
    
Gamma correction is given by: 
\begin{equation}
    I'=c\cdot I^{\gamma}, 
\end{equation}
where $I$ represents the original image, $I'$ is the transformed image, $c$ is a constant, and $\gamma$ is the gamma value.
    
% \begin{itemize}
%     \item When $\gamma > 1$, the dark areas of the image become brighter, enhancing shadow details.
%     \item When  $0 < \gamma < 1$, the bright areas of the image become darker, enhancing highlight details.
%     \item When $\gamma < 0$, the image exhibits a negative effect (inverted colors).
% \end{itemize}

Gamma correction allows better identification of tissues and details by enriching the contrast of fundus images.
In \cite{somasundaram2011medical}, an automatic gamma correction method was introduced based on the cumulative histogram to conveniently enhance image contrast.
By combining adaptive gamma correction and homomorphic filtering, Tiwari et al. \cite{tiwari2016brightness} proposed an adaptive gamma correction method to enhance the contrast of medical images while maintaining their brightness. 
Another adaptive gamma correction technique utilizing weighted histogram equalization was developed in \cite{agarwal2017medical}, surpassing traditional histogram equalization in terms of visual quality and maximum entropy preservation.
Acharya et al. \cite{acharya2024directed} presented a texture-adaptive gamma correction method to optimize texture recognition and enhancement rates, resulting in sharper edges and richer textures.

\subsubsection{Image Filtering}

\begin{figure}[tbp]
        \begin{centering}
            \includegraphics[width=0.9\linewidth]{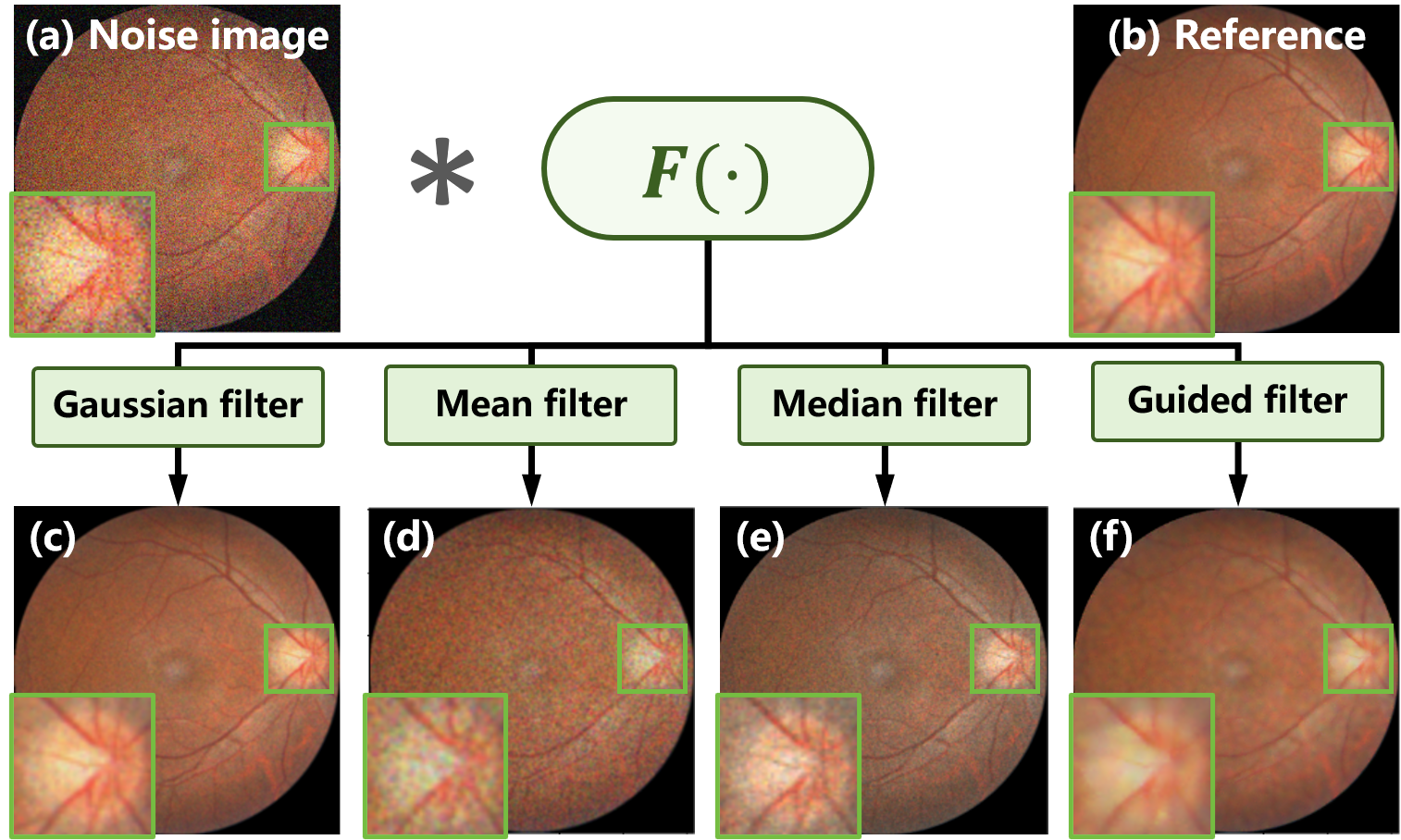}
            \par
        \end{centering}
        %\vspace{-2.5ex}
    \caption{Image filter exhibition. (a) a noise image degraded by Gaussian noise, (b) the original high-quality image for reference, (c) Gaussian filter outcome, (d) Mean filter outcome, (e) Median filter outcome, and (d) Guided filter outcome.}
    \vskip -5pt
    \label{fig:enhance-4}
    \end{figure}

Image filtering entails adjusting the pixel values of an image with a kernel (a small matrix of coefficients) according to the values of nearby pixels, allowing for a range of enhancement effects. 
As illuminated in Fig.~\ref{fig:enhance-4}, the filter's effect is determined by the kernel, which convolves with the image, sliding over it to calculate the weighted sum of pixel values under the kernel at each position. The filtered image is obtained by replacing the original value of the central pixel with the calculated sum.

The general formula for image filtering is expressed as:
\begin{equation}
I'(x, y) = \sum_{i=-a}^{a} \sum_{j=-b}^{b} h(i, j) \cdot I(x-i, y-j)
\end{equation}
where $I(x, y)$ and $I'(x, y)$ are the pixel values of the input and filtered image at position (x, y). 
$h(i, j)$ represents the weight assigned to the filter kernel at position $(i, j)$, defining the type of image filter, with the kernel size being determined by $a$ and $b$.

The common image filters in IQE include:
\begin{itemize}
    \item \textbf{Gaussian Filtering} is a linear smoothing filter, which applies a Gaussian kernel to individual pixels, performing a weighted average with neighboring pixels to remove high-frequency noise while preserving the main features of the image, as shown in Fig.~\ref{fig:enhance-4} (c). It is suitable for eliminating Gaussian noise.
    
    \item \textbf{Mean Filtering} employs a window surrounding each pixel, replacing the central pixel with the average of all pixels within the window to achieve smoothing, as shown in Fig.~\ref{fig:enhance-4} (d). 
    It is suitable for removing random noise with a uniform distribution.

    \item \textbf{Median Filtering} is a non-linear filtering method where each pixel's value is replaced by the median value of its neighboring pixels. This filtering efficiently eliminates salt-and-pepper noise while conserving edge details, as shown in Fig.~\ref{fig:enhance-4} (e).
    
    \item \textbf{Guided Filtering} employs a guiding image, which can be the input image itself or another image, to direct the filtering process, enhancing image smoothness while maintaining details and edges. This filtering adeptly balances noise reduction with detail preservation, as shown in Fig.~\ref{fig:enhance-4} (f).
\end{itemize}

In \cite{zhao2017automatic}, symmetric image filters were employed to estimate the local phase unaffected by illumination changes, enhancing local structures in fundus images.
Cao et al. \cite{cao2020retinal} utilized low-pass filtering to enhance the contrast of fundus images, eliminating low-frequency components in the root domain.
An adaptive weighted mean filter was developed in \cite{yugander2020mr} to enhance image contrast.
Anoop et al. \cite{anoop2022medical} introduced a bilateral image filter to remove noise while preserving edges.
Gaussian filters were leveraged in \cite{li2022structure} and \cite{li2022annotation} to decompose and preserve fundus structures in the enhancement models.

\subsubsection{Statistical Prior Techniques}
In IQE, statistical prior techniques generally refer to leveraging known information derived from data distributions, physical models, or prior knowledge to guide the enhancement process. These priors are crucial in refining image quality, enhancing visual effects, and boosting processing efficiency. However, identifying and incorporating statistical attributes remains a formidable task. Currently, the prominent techniques applied in fundus IQE are green channel and dark channel priors.
    
\noindent\textbf{Green Channel Prior}
 
\begin{figure}[htbp]
        \begin{centering}
            \includegraphics[width=\linewidth]{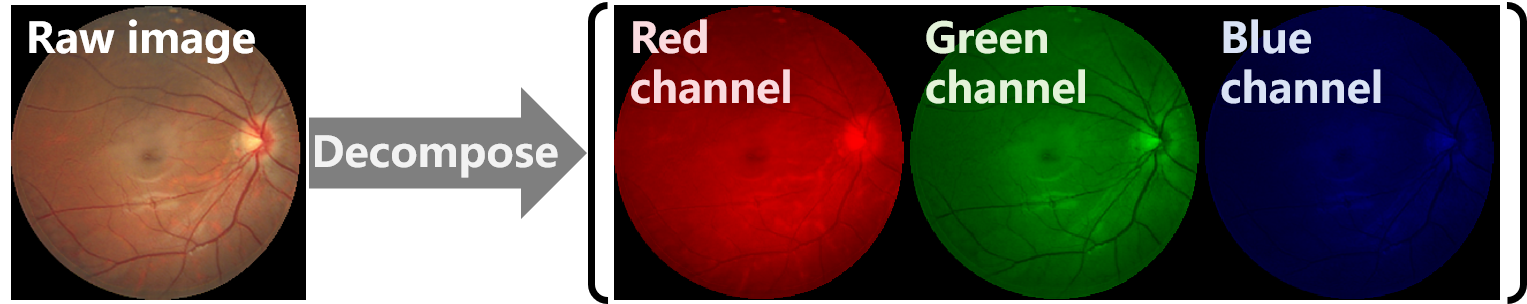}
            \par
        \end{centering}
        %\vspace{-2.5ex}
    \caption{The green channel highlights the fundus vascular structure.}
    \vskip -5pt
    \label{fig:enhance-6}
    \end{figure}

The green channel prior in fundus image processing refers to the observation that the green channel of a color fundus image often provides the best contrast and clarity for visualizing retinal vasculature and other important anatomical structures~\cite{guo2021joint}, as demonstrated in Fig.~\ref{fig:enhance-6}.
This is because blood absorbs green light more strongly than red or blue light, resulting in darker blood vessels in the green channel, making them stand out more clearly against the background.

Based on the green channel prior, heuristic IQE algorithms have been proposed that leverage the inherent contrast and clarity of the green channel to improve the fundus image quality. 

In \cite{zhou2017color}, the RGB channels were corrected by rectifying the value channel in the HSV color space in conjunction with CLAHE.
The green channel prior and cubic function were integrated in \cite{cao2021detail} to improve the clarity of fundus images.
Zhang et al. \cite{zhang2022double} employed green channel as the primary transmission map and adjusted it using a revised cubic function to mitigate excessive enhancement.
Priyadharsini et al. \cite{priyadharsini2023retinal} leveraged the color dominance within images to assess information distribution among the RGB channels and conducted enhancements in the color space to refine overall brightness and contrast.

\noindent\textbf{Dark Channel Prior}

The dark channel prior is a powerful observation originally identified for single-image dehazing~\cite{he2010single}.
It states that in a clear image, most objects reflect some light in at least one color channel, resulting in low values in the dark channel.
This principle facilitates the estimation of the transmission map in hazy images, illustrating how light permeates through the haze to the camera, ultimately allowing the reconstruction of a haze-free image through the inversion of the atmospheric scattering model.

Given that fundus images affected by media opacity share a comparable imaging model with hazy images, the dark channel prior has been integrated into fundus IQE algorithms. 
A haze image $I$ from natural environments is formulated as: 
\begin{equation}
    I(x,y)=J(x,y)\cdot t(x,y)+A\cdot[1-t(x,y)],
    \label{dark}
\end{equation}
where $(x,y)$ represents the pixel position, $J$ is the scene content (the desired haze-free image), $A$ is the atmospheric light, and $t$ is the transmission function describing the fraction of light that reaches the camera.

% \begin{figure}[htbp]
%     \begin{centering}
%         \includegraphics[width=\linewidth]{images/enhance/4-2-7.png}
%         \par
%     \end{centering}
%     %\vspace{-2.5ex}
% \caption{Imaging model of fundus photography for cataract patients.}
% \vskip -5pt
% \label{fig:enhance-7}
% \end{figure}

For cataract patients, the fundus image $I$ is formulated following~\cite{Peli_Peli_1989} as:
% (refer to Fig.~\ref{fig:enhance-7}): 
\begin{equation}
        I(x,y)=\alpha \cdot L \cdot \gamma(x,y) \cdot t(x,y) + L(1-t(x,y)),
        \label{cataract}
\end{equation}
where $\alpha$ is the attenuation constant of retinal illumination, $L$ denotes the camera illumination, and $\gamma$ and $t$ represent the retinal reflection function and lens transmission function, respectively.

As Eq.~\ref{cataract} can be simplified to Eq.~\ref{dark}, the dark channel prior is introduced into fundus IQE.
Cheng et al. \cite{cheng2018structure} introduced an IQE method for fundus images that combines a simplified dark channel prior with edge-preserving smoothing to estimate transmission maps, reducing haze and scattering while preserving structural details.
Adapting the dark channel prior to medical contexts, Singh et al. \cite{singh2019dark} estimated transmission maps and atmospheric light by restoring scene radiance through normalization and soft matting, and applying gamma correction to improve brightness and contrast.
Drawing on the dark channel prior and the double-pass fundus reflection model, Zhang et al. \cite{MUTE} proposed a multilevel stimulated denoising strategy to dehaze single cataractous retinal images by converting the estimation of cataract transmission function into an image denoising problem.

\subsection{Learning-based IQE Methods}
With the progress in deep learning, learning-based IQE algorithms have rapidly developed in fundus imaging, enabling the automatic learning of the mapping from low-quality data to high-quality data.
% As depicted in Fig.~\ref{fig:enhance-8}, t
These learning-based IQE algorithms, often employing image-to-image (I2I) translation generators, learn the mapping from low-quality to high-quality images using paired or unpaired funds image data

    % \begin{figure}[htbp]
    %     \begin{centering}
    %         \includegraphics[width=\linewidth]{images/enhance/4-2-8.png}
    %         \par
    %     \end{centering}
    %     %\vspace{-2.5ex}
    % \caption{Learning-based IQE paradigm based on I2I translation generator.}
    % \vskip -5pt
    % \label{fig:enhance-8}
    % \end{figure}

\subsubsection{IQE Models Trained on Paired Data}
Similar to existing FR-IQA algorithms, learning-based IQE models relying on paired data are trained with low- and high-quality paired fundus images.

Denote an I2I generator as $G_{\theta}$, where $\theta$ is the parameters of the generator model. 
The vanilla objective of $G_{\theta}$ for IQE using paried data is given by:

\begin{equation}
    \underset{\theta}{arg\,min}\left \|G_{\theta}(I)-R\right \|_1,
    \label{Eq:IQE}
\end{equation}
where $I$ is the low-quality image, $R$ denotes the paired high-quality reference.

\begin{figure}[tbp]
    \begin{centering}
        \includegraphics[width=\linewidth]{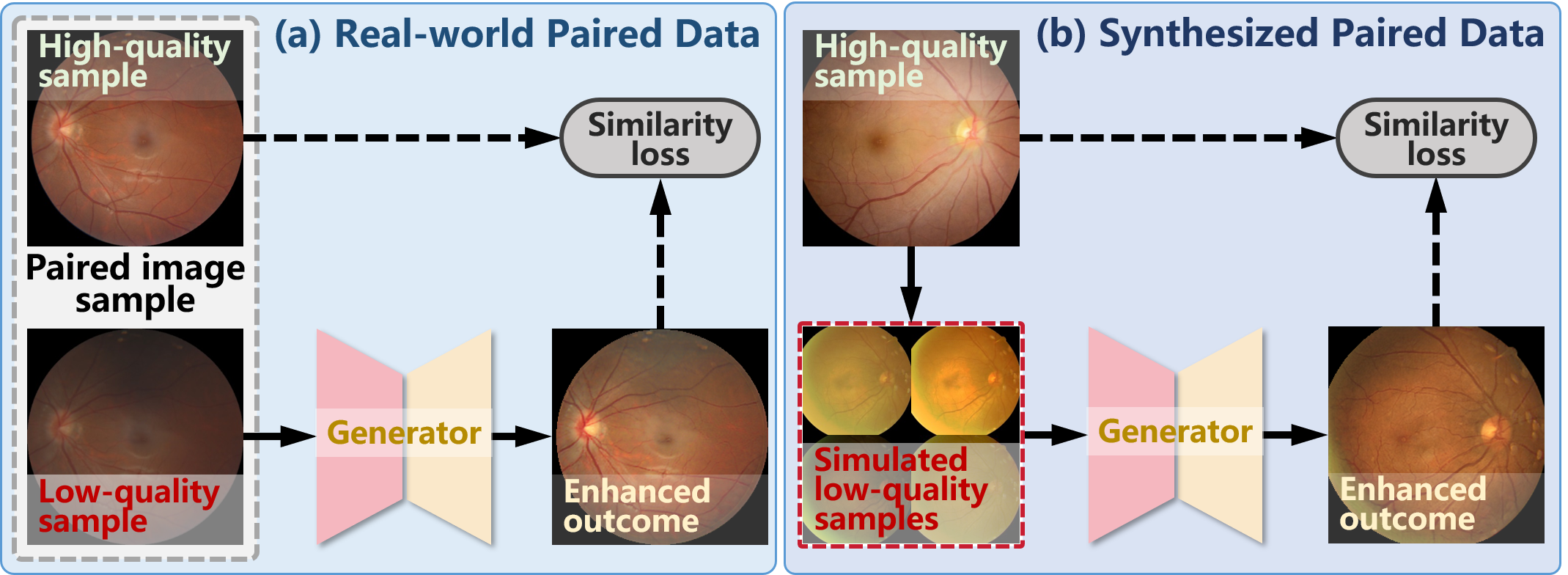}
        \par
    \end{centering}
    %\vspace{-2.5ex}
\caption{Learning-based IQE architectures using real-world and synthesized paired data.}
\vskip -5pt
\label{fig:enhance-9}
\end{figure}

While paired data offer a straightforward approach for training IQE models, collecting paired fundus image data is challenging and even impractical in real-world scenarios.
Consequently, as demonstrated in Fig.~\ref{fig:enhance-9}, there exist two categories of paired data in training learning-based IQE models: 1) real-world paired data collected via repeated capture and 2) synthesized paired data obtained by simulating low-quality fundus images from high-quality ones.

\noindent\textbf{Real-world Paired Data}

Real-world data accurately reflects noise, blur, and distortion issues present in specific clinical environments, allowing IQE models to effectively learn parameters for enhancing corresponding low-quality fundus images.
However, collecting paired data in the real-world is challenging, especially in fields like medicine.

Specifically, $G_{\theta}$ necessitates pixel-level pairs of -quality fundus images, typically obtained through repeated image capture.
However, due to the uncontrolled relative eye motion in imaging, repeat captured images are susceptible to replacement, resulting in pixel-level misalignments. 
Therefore, the collection of paired data and the acquisition of pixel-level annotations are both costly and time-intensive.
Additionally, obtaining high-quality fundus images through repeat capture is unfeasible for certain individuals such as infants and patients with cataracts.

Despite the challenges of repeated data capture and feasibility issues in specific scenarios, real-world paired data usually delivers satisfactory training performance for IQE models, as shown in Fig.~\ref{fig:enhance-9} (a).
Given a dataset of real-world pairs, $\{I_k, R_k\}$, the optimization objective of $G_{\theta}$ is consistent to Eq.~\ref{Eq:IQE}:

\begin{equation}
    \underset{\theta}{arg\,min}\textstyle \sum_{k=1}^K|| G_{\theta}(I_k)-R_k||_1,
\end{equation} 
where $k$ is the index of pairs in the dataset, and $K$ denotes the dataset volume.

The limitations in acquiring low- and high-quality paired data from real-world scenarios posed significant challenges to the development of the learning-based fundus IQE model. 
To alleviate the data collection limitations, Upadhyay et al. \cite{upadhyay2019mixed} introduced a mixed-supervision Generative Adversarial Network (GAN) that leverages training data across different quality levels, including high and medium quality.
By utilizing 120 image pairs acquired through repeated captures, Deng et al. \cite{Deng_Cai_Chen_Gong_Bao_Yao_Fang_Zhang_Ma} presented a Transformer-based GAN designed to enhance fine details and anatomical structures, resulting in remarkable improvements in tasks like vessel segmentation and optic disc detection.

\noindent\textbf{Synthesized Paired Data}

Although real-world paired data provide accurate degradation information, the challenges associated with collection and annotation constrain their practicality.
To overcome this limitation, synthesized paired data are commonly employed for model training.
Accordingly, low-quality fundus images are synthesized from high-quality ones to create the image pairs.
Moreover, as illuminated in Fig.~\ref{fig:enhance-9} (b), to boost training robustness, a range of degradations is generally simulated to synthesize diverse low-quality images from a single high-quality image.

Denote a high-quality fundus image dataset as $\{R_k\}$. The low-quality image synthesized from $R_k$  is represented by $\tilde{I}_{kn}$, with $n$ indexing the degraded images.
Consequently, the paired training dataset is expressed as $\{\tilde{I}_{kn}, R_k\}$, and the optimization objective of $G_{\theta}$ is defined as:
\begin{equation}
    \underset{\theta}{arg\,min}\textstyle \sum_{k=1}^k\textstyle \sum_{n=1}^N|| G_{\theta}(\tilde{I}_{kn})-R_k||_1,
\end{equation} 
where $N$ refers to the number of degraded images.

To overcome the challenges associated with collecting paired real-world data, various simulation models have been proposed for synthesizing pairs of low- and high-quality fundus images.
By examining the imaging system, Shen et al. \cite{shen2020modeling} introduced degradation models for light transmission disturbance, image blurring, and retinal artifact in fundus photography, facilitating the training of fundus IQE models. 
Luo et al. \cite{luo2020dehaze} introduced a GAN-based technique to synthesize cataract-like images for training restoration models of cataractous fundus images.
The degradation model for cataracts was suggested in \cite{li2022annotation} to develop restoration models in the absence of corresponding high-quality references.
Incorporating a synthetic forward model of degradation with a direct diffusion bridge, Kim et al. \cite{kim2024fundus} introduced a diffusion-based image enhancement network for fundus images.

Furthermore, researchers have identified that the inherent disparity between the synthesized and real-world data impacts the effectiveness of enhancement during inference. Substantial endeavors have been undertaken to bridge this gap. Li et al. \cite{li2022annotation} and Guo et al. \cite{guo2023bridging} integrate the synthesized paired data and real low-quality data to implement domain adaptation and alleviate this discrepancy. 
Domain generalization techniques were introduced in \cite{li2022structure} and \cite{Agenericfundusimageenhancementnetwork} to construct robust enhancement models using a mix of synthesized paired data, allowing these models to be generalized to unseen degraded images.

\subsubsection{IQE Models Trained on Unpaired Data.}
Considering the expenses and feasibility associated with collecting paired fundus image data, efforts have been directed to train IQE models using unpaired data.
By circumventing the necessity of obtaining images of varying quality from identical individuals, unpaired data significantly alleviate the complexities of data collection, thereby expanding the spectrum of image scenarios and conditions and enriching the pool of available data samples.

\begin{figure}[tbp]
    \begin{centering}
        \includegraphics[width=0.9\linewidth]{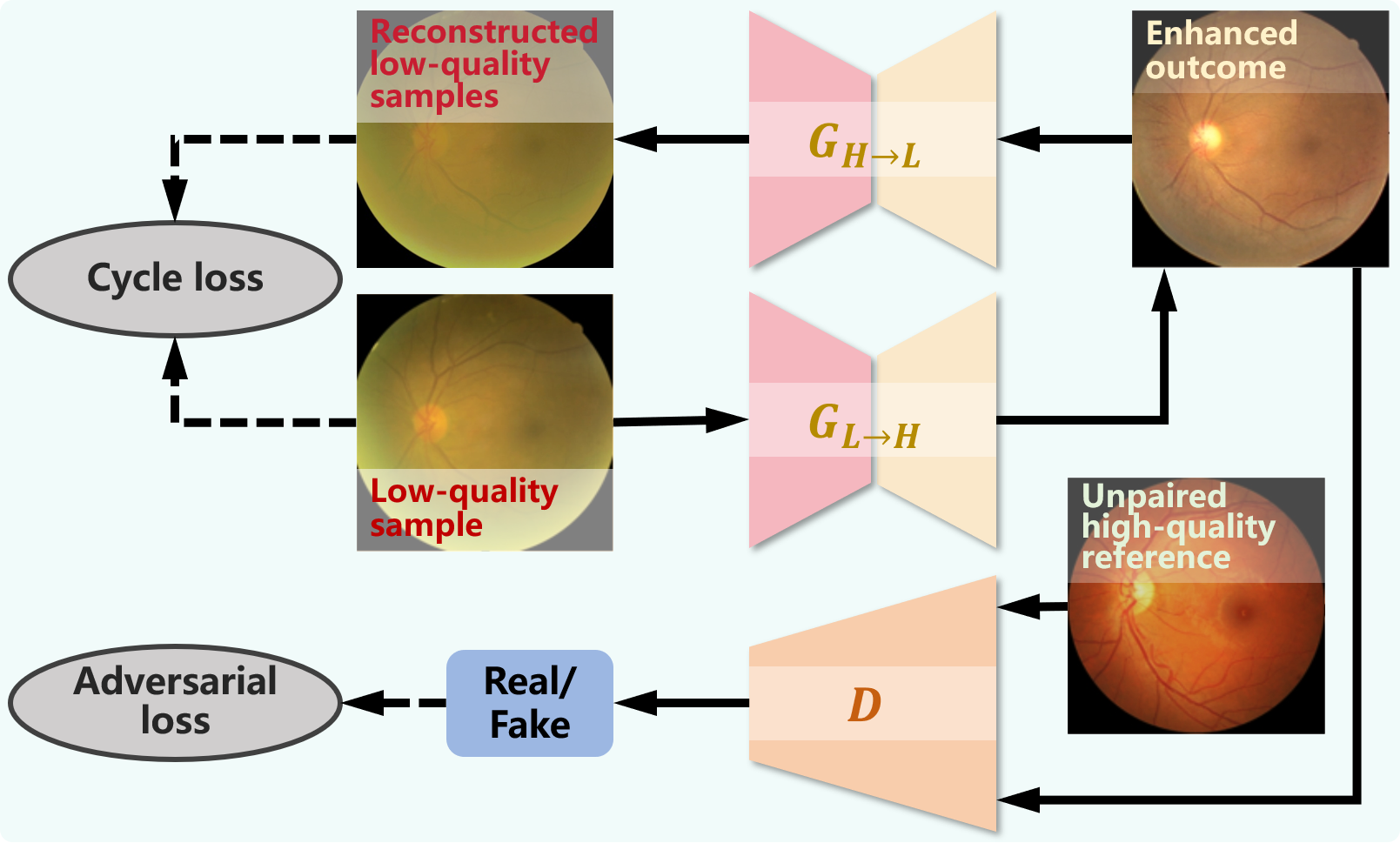}
        \par
    \end{centering}
    %\vspace{-2.5ex}
\caption{Learning-based IQE architectures using unpaired data.}
\vskip -5pt
\label{fig:enhance-10}
\end{figure}

Motivated by the advancements in unpaired I2I translation, IQE models based on unpaired data are progressing rapidly.
Illustrated in Fig.~\ref{fig:enhance-10}, CycleGAN~\cite{zhu2017unpaired} emerges as the prevalent architecture employed in fundus IQE.
Instead of image pairs from identical individuals, CycleGAN leverages a cycle-consistency loss to enable the training of I2I generators bridging the data domains of low- and high-quality fundus images.

\begin{table*}[!t]
\centering
\footnotesize
\caption{Representative fundus IQE models. The background color indicates the model is based on \colorbox{gray1}{paired} or \colorbox{gray2}{unpaired} data.}
\label{tab:iqe}
\begin{tabular}{l|l|l|l}
\hline
Article &  Contribution & Quality-related Dataset & Downstream Task \\
\hline
\rowcolor[gray]{.95} \cite{shen2020modeling} &  Degradation models \& structure constraint & EyeQ & Segmentation \\
\hline
\rowcolor[gray]{.95} \cite{luo2020dehaze} &  Low-quality image synthesis model & / & / \\
\hline
\rowcolor[gray]{.95} \cite{Deng_Cai_Chen_Gong_Bao_Yao_Fang_Zhang_Ma} & Real-world paired data \& self-attention block & RF & Segmentation \\
\hline
\rowcolor[gray]{.95} \cite{li2022annotation} & Cataract model \& domain adaptation & RCF \& iSee & Diagnosis \\
\hline
\rowcolor[gray]{.95} \cite{Agenericfundusimageenhancementnetwork} &  Structure constraint \& domain generalization & FIQ \& RCF \& EyeQ \& iSee & Segmentation \& diagnosis \\
\hline
\rowcolor[gray]{.95} \cite{guo2023bridging} & Structure constraint \& domain adaptation  & EyeQ & Segmentation \& diagnosis \\
\hline
\rowcolor[gray]{.95} \cite{kim2024fundus} & Diffusion-based IQE  & EyeQ \& FPE & Segmentation \\
\hline
\rowcolor[gray]{.85} \cite{ma2021structure} & CycleGAN with structure constraint & iSee & Localization \\
\hline
\rowcolor[gray]{.85} \cite{cheng2021prior} & CUT with fundus prior & EyeQ & / \\
\hline
\rowcolor[gray]{.85} \cite{cheng2021secret} & CUT with importance-guidance & EyeQ & Segmentation \\
\hline
\rowcolor[gray]{.85} \cite{yang2023retinal} & CycleGAN with multi-task constraint & EyeQ & Segmentation \& tracking\\
\hline
\rowcolor[gray]{.85} \cite{he2023hqg} & CycleGAN with high-quality guidance & iSee & / \\
\hline
\rowcolor[gray]{.85} \cite{hou2024reference} & IQA guided domain adaptation & EyeQ & / \\
\hline
\rowcolor[gray]{.85} \cite{EnhancingandAdapting} & Source-free domain adaptation & FIQ \& RCF \& RF \& iSee & Segmentation \& diagnosis \\
\hline
\end{tabular}
\end{table*}

The unpaired data consist of a low-quality subset $\{I_k\}$ and a high-quality subset $\{R_n\}$, where $\{I_k\}$ and $\{R_n\}$ are mutually independent.
Within the architecture are two I2I generators: one ($G_{L\to H}$) for translating low-quality samples to the high-quality domain and the other ($G_{H\to L}$) for the reverse translation.
Additionally, an adversarial loss $\mathcal{L} _{adv}$ is introduced by a discriminator $D$ to enable the learning of mappings between the domains of $\{I_k\}$ and $\{R_n\}$ with the generators.
$\mathcal{L} _{adv}$ is given by:

\begin{equation}
\begin{aligned}
    \mathcal{L} _{adv}(G_{L\to H}, D)& = \mathbb{E}[\log D(R_n)]\\
    &+\mathbb{E}[\log (1-D(G_{L\to H}(I_k)))],
\end{aligned}
\end{equation}

where $G_{L\to H}$ endeavors to translate $I_k$ to a enhanced outcome $G_{L\to H}(I_k)$, while $D$ strives to distinguish between $G_{L\to H}(I_k)$ and real high-quality samples $R_n$.
Notably, a corresponding adversarial loss for $G_{H\to L}$ is also utilized with another discriminator, but it is omitted in this section.

To maintain content fidelity in $G_{L\to H}(I_k)$, a cycle-consistent loss $\mathcal{L} {cyc}$ is integrated, following the notion that for each image $I_k$, the image translation cycle should be capable of returning it to the original image, i.e. $ I_k\Rightarrow G{L\to H}(I_k) \Rightarrow G_{H\to L}(G_{L\to H}(I_k)) \approx I_k$.
Consequently, $\mathcal{L} _{cyc}$ is expressed as:

\begin{equation}
\begin{aligned}
    \mathcal{L} _{cyc}(G_{L\to H},G_{H\to L}) &= \mathbb{E}\left \| G_{H\to L}(G_{L\to H}(I_k))-I_k \right \| _1\\
    & + \mathbb{E}\left \| G_{L\to H}(G_{H\to L}(R_n))-R_n \right \| _1
\end{aligned}
\end{equation}

By combining $\mathcal{L} _{adv}$ and $\mathcal{L} _{cyc}$, it becomes feasible to train IQE models using unpaired data.
Nonetheless, due to the dissimilar content within low- and high-quality images, generation artifacts may be introduced in the outcomes by the enhancement model~\cite{li2022sample}.

Multiple unpaired IQE models have been developed based on CycleGAN, incorporating additional constraints to ensure content fidelity.
Zhang et al. \cite{zhao2019data} introduced a dynamic retinal image feature to restrict CycleGAN, aiming to prevent excessive enhancement of severely blurred regions.
% \cite{Park_Efros_Zhang_Zhu_2020} utilized mutual information to enhance structural consistency between raw and enhanced fundus images.
Similarly, illumination and structural regularization were proposed in \cite{ma2020cycle} and \cite{ma2021structure} to improve fundus image illumination uniformity and enhance structural details.
Yang et al. \cite{yang2023retinal} introduced a high-frequency prior to mitigate artifacts in the enhancement of fundus vessels.
A variational framework and a bi-level learning scheme have been proposed in \cite{he2023hqg} to enhance both visual quality and downstream task performance.
Hou et al. \cite{hou2024reference} introduced fundus IQA models to guide the enhancement using only low-quality images.

On the other hand, the unpaired I2I translation architecture of CUT~\cite{park2020contrastive} has also been utilized in fundus IQE.
Prior knowledge of fundus images has been integrated with CUT in \cite{cheng2021prior} and \cite{cheng2021secret} to enable enhancement in the absence of paired data.

\subsection{Challenges and Insight}
A summary of the representative fundus IQE models is outlined in Table~\ref{tab:iqe}.
The manually crafted features in handcrafted IQE methods offer a strong interpretability for the enhancement procedure and outcomes.
However, these features are often tailored to specific degradations, resulting in limited versatility across different types of degradations. 
Moreover, the handcrafted methods struggle to implement intricate nonlinear mappings, resulting in subpar enhancement performance when faced with complex degradation scenarios.
In contrast, learning-based IQE methods can automatically learn features and construct nonlinear mappings for diverse degradation types from the training data without manual intervention.
Nonetheless, the underlying behavior of learning-based methods is often challenging to interpret, hindering their application in medical scenarios where a high degree of interpretability is required.
To summarize, the trend in fundus IQE is shifting towards incorporating manually crafted features in learning-based IQE methods. This trend allows for the convenient construction of IQE models and provides adequate interpretability to comprehend the process and outcomes effectively.

\begin{figure}[htbp]
    \begin{centering}
        \includegraphics[width=0.8\linewidth]{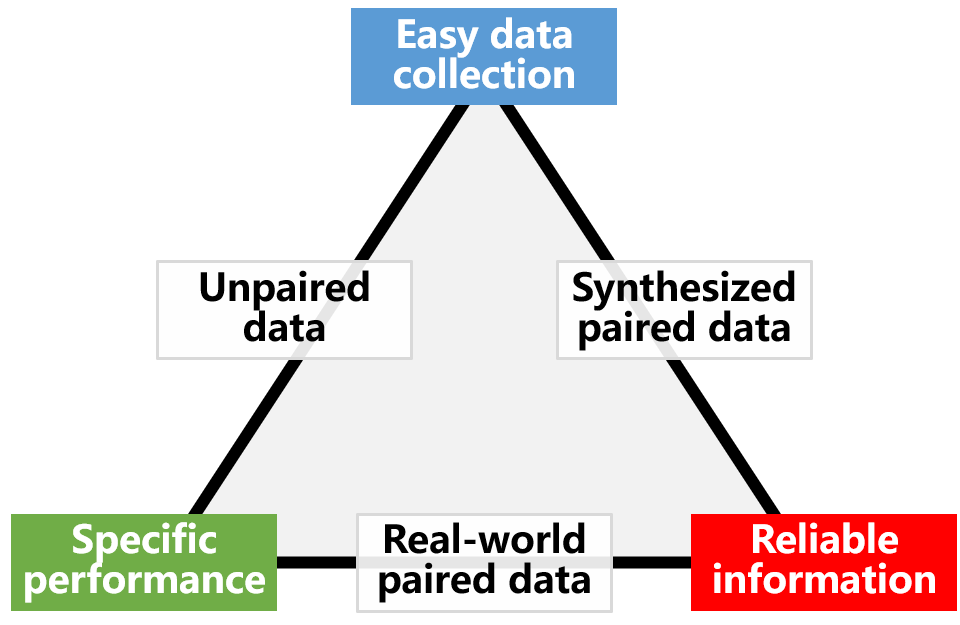}
        \par
    \end{centering}
    %\vspace{-2.5ex}
\caption{Capacity attributes of various training data types.}
\vskip -5pt
\label{fig:enhance-11}
\end{figure}

On the other hand, the collection of training data poses a significant challenge in the advancement of fundus IQE methods. 
While the methods based on synthesized paired data and unpaired data circumvent the challenges associated with collecting paired data from the real-world, certain limitations persist.
The attributes of these data types are demonstrated in Fig.~\ref{fig:enhance-11}.
Real-world paired data accurately reflects the degradation traits and data distribution, leading to superior enhancement performance specific to the training data.
And pixel-level paired training data ensures reliable information preservation of information in the enhanced outcomes.
The inherent disparities between synthesized data and real-world data cannot be completely eliminated, potentially resulting in a performance drop when applying a model trained on synthesized data for real-world data inference.
Although unpaired data are easier to collect in real-world settings, the absence of paired references hinders the model's ability to preserve crucial content, consequently leading to information distortion in the outcomes.

\section{Fundus IQA and IQE in Practical Applications}
In addition to investigating the cutting-edge advancements in IQA and IQE techniques for fundus images, this chapter delves into the practical applications of the existing methods. Specifically, two significant folds are discussed. Firstly, the challenges in deploying and applying IQA and IQE techniques are discussed, along with the strategies to overcome these challenges. Secondly, the paradigms to incorporate IQA and IQE techniques into ophthalmology workflows to enhance fundus image understanding and diagnosis. 

\subsection{Deployment challenge in fundus IQA and IQE}
Similar to most medical scenarios, a fundamental challenge in deploying IQA and IQE algorithms for fundus images is the availability of data. This availability, in turn, gives rise to challenges related to generalizability and interpretability.

\subsubsection{Data availability}
Gathering a sufficient amount of annotated data for training models has long been a persistent issue in the field of medical image research. Fundus images, comparatively, offer a convenient and cost-effective medical imaging modality, facilitating data collection. However, the challenge posed by annotations exacerbates the data scarcity in the context of IQA and IQE for fundus images.

\begin{table*}[htbp]
\centering
\begin{threeparttable}
\footnotesize
\caption{Commonly used datasets in fundus IQA and IQE studies.}
\label{tab:datasets}
\begin{tabular}{l | r | c | l | l}
\hline
Dataset & Instance \# & Quality \# & Task labels & Data description \\
\hline
\rowcolor[gray]{.90} EyeQ~\cite{fu2019evaluation}& 28792 & 3 & DR grading & DR grading dataset with image quality labels \\
\hline
\rowcolor[gray]{.90} DRIMDB~\cite{csevik2014identification} & 216 & 2 & -- & DR images with quality labels\\
\hline
\rowcolor[gray]{.90} FIVES~\cite{jin2022fives} & 800 & 2 & Vessel Segmentation & Annotations for quality and segmentation\\
\hline
\rowcolor[gray]{.90} MSHF~\cite{jin2023mshf} & 1302 & 2 & Disease diagnosis & Annotations for quality, degradation, and diagnosis\\
\hline
\rowcolor[gray]{.90} FPE~\cite{kim2024fundus} & 1202 & 2 & -- & 50 bad-quality samples VS 1152 high-quality ones\\
\hline
\rowcolor[gray]{.90} ODIR \tnote{1} & 10000 & 2 & Disease diagnosis & Diagnosis in the presence and absence of noise\\
\hline
\rowcolor[gray]{.90} DeepDRiD~\cite{LIU2022100512} & 2000 & 2 & --& Fundus photography and ultra-widefield fundus imaging\\
\hline
\rowcolor[gray]{.90} iSee \tnote{2} & 10000 & 2 & Disease diagnosis & Diagnosis for patients with and without cataracts\\
\hline
\rowcolor[gray]{.90} HRF~\cite{kohler2013automatic} & 18 $\times$ 2 & 2 & -- & Low- and high-quality paired data\\
\hline
\rowcolor[gray]{.90} RF~\cite{Deng_Cai_Chen_Gong_Bao_Yao_Fang_Zhang_Ma} & 120 $\times$ 2 & 2 & -- & Low- and high-quality paired data\\
\hline
\rowcolor[gray]{.90} RCF~\cite{li2022annotation} & 26 $\times$ 2 & 2 & -- & Fundus image pairs from pre- and post-cataract surgery\\
\hline
\rowcolor[gray]{.90} FIQ~\cite{liu2022degradation} & 196 $\times$ 2 & 2 & -- & Low- and high-quality paired data\\

\hline
\end{tabular}
% \begin{tablenotes}
% \scriptsize
1 https://odir2019.grand-challenge.org/ 2 https://www.imed-lab.com/
% \item[3] https://github.com/HzFu/EyeQ
% \item[4] https://www.nature.com/articles/s41597-023-02188-x
% \item[6] https://www.nature.com/articles/s41597-022-01564-3\#Sec1
% \item[5] www5.cs.fau.de/research/data/fundus-images/
% \end{tablenotes}
\end{threeparttable}
\end{table*}

\begin{figure}[htbp]
    \begin{centering}
        \includegraphics[width=\linewidth]{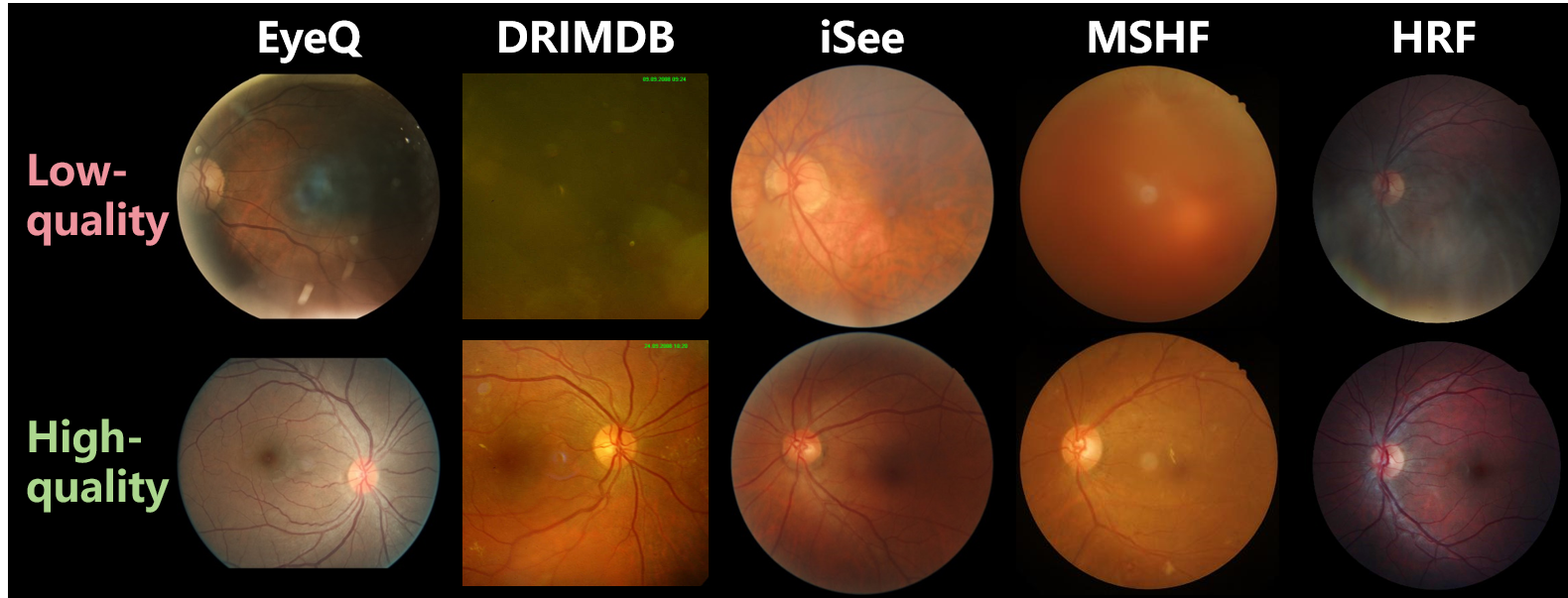}
        \par
    \end{centering}
\caption{Examples of variance in data distribution and annotation across fundus datasets.}
\vskip -5pt
\label{fig:application_domain_shift}
\end{figure}

\noindent\textbf{Research Data}

We offer an overview of frequently employed datasets in IQA and IQE research for fundus images, as outlined in Table~\ref{tab:datasets}. 
These datasets are typically applicable to both IQA and IQE research.
The attributes of these datasets, such as quality grading levels, collection conditions, and the presence or absence of paired data, are summarized.

Specifically, EyeQ~\cite{fu2019evaluation} categorizes fundus image quality into three levels: "Accept", "Usable", and "Reject". In contrast, other datasets grade fundus images using only two quality levels, namely "High (Good) quality"  and "Low (Bad) quality".
It is important to note that quality annotations tend to be distinct among datasets, as exhibited in Fig.~\ref{fig:application_domain_shift}. In other words, there are variations in how high and low quality are defined across datasets, and quality labels may not be interchangeable between different datasets in developing IQA algorithms.
Additionally, MSHF~\cite{jin2023mshf} and a portion of ODIR contain more granular quality annotations, encompassing evaluations of factors like illumination, sharpness, and contrast in fundus images.

On the other hand, low- and high-quality unpaired data are more common in datasets, which can only be utilized to develop unpaired IQE algorithms or synthesize paired data to develop paired algorithms. 
Low- and high-quality paired data are available in HRF, RF~\cite{Deng_Cai_Chen_Gong_Bao_Yao_Fang_Zhang_Ma}, RCF~\cite{li2022annotation}, and FIQ~\cite{liu2022degradation}, enabling direct use in developing algorithms for paired IQE.

Furthermore, some datasets include annotations for downstream tasks such as DR grading, disease diagnosis, and structural segmentation, expanding the utility of these datasets.

\noindent\textbf{Data Collection}

Although we have provided a summary of the available fundus image datasets for IQA and IQE research, the collection of data and annotations remains an obstacle to algorithm application and deployment.

For instance, in certain scenarios where IQE is particularly needed, it is precisely unfeasible to obtain high-quality fundus images as the annotation of low-quality ones for model training, thereby stalling algorithm development.
As depicted in Fig~\ref{fig:application_datascarcity}, since it is difficult to capture high-quality fundus images from infants and cataract patients, IQE algorithms are needed but there is also a lack of corresponding paired training data.

Alternatively, synthesizing paired data is a convenient strategy for collecting the annotated training data for fundus IQE models.
Degradation models proposed in \cite{shen2020modeling} and \cite{li2022annotation} enable the generation of low-quality images from high-quality ones, and the synthesized paired data is leveraged to train IQE models. 

Notably, neither public datasets nor synthesized data can be guaranteed to be completely consistent with the data encountered in the inference phase. 
Consequently, the presence of low- and high-quality paired data from real-world applications is indispensable for the deployment of IQE algorithms.
Because high-quality data references collected from relevant clinical scenarios are always the gold standard for evaluating the qualification of IQE performance.

\subsubsection{Generalizability}
Due to the difficulties involved in collecting fundus data and annotations, researchers often rely on the public data summarized in Table~\ref{tab:datasets} and synthesized data to craft IQE algorithms. 
However, in practical scenarios, changes in data sources lead to variations in data distribution and annotations (as shown in Fig.~\ref{fig:application_domain_shift}), and discrepancies between synthesized and raw data are also inevitable.
These issues refer to domain shifts, where a model trained on the source domain may experience performance degradation when applied to the target domain. 
Therefore, recent studies have introduced domain adaptation and domain generalization techniques to mitigate the impacts of domain shifts.

\noindent\textbf{Domain Adaptation}

Domain adaptation involves adapting a model trained on a source domain to perform well on a different but related target domain.

In the realm of fundus IQA, Shen et al. \cite{shen2020domain} introduced adversarial learning to align features across IQA datasets through discriminator confusion.
Lin et al. \cite{lin2023domain} proposed a teacher-student framework in fundus IQA for global feature alignment and fine-grained learning for domain shifts between different IQA datasets. 

For fundus IQE, Li et al. \cite{li2021Restoration} introduced an unsupervised restoration method for cataract images using domain adaptation to extend the enhancement model's applicability from synthetic to real data.
Li et al. \cite{li2022annotation} proposed ArcNet, which leverages knowledge transfer from synthesized to real cataract data by minimizing the distinction between enhanced images.
Guo et al. \cite{Bridging} devised an end-to-end optimized teacher-student framework to simultaneously conduct image enhancement and domain adaptation, transferring insights from synthetic datasets to real-world scenarios.
Guo et al. \cite{guo2024multi} devised a dynamic filtering network to adaptively enhance fundus images by identifying degradation patterns and generating tailored filters accordingly.
Moreover, Li et al. \cite{EnhancingandAdapting} developed a source-free domain adaptation algorithm for fundus IQE that harnesses the value of post-deployment data to adapt and optimize the enhancement model during the inference stage.

\noindent\textbf{Domain Generalization}

Domain generalization trains a model on multiple source domains with different data distributions to excel on unseen target domains.

Liu et al. \cite{liu2022Domain} pioneered the integration of domain generalization into fundus IQE by aligning frequency features extracted from images with diverse degradations.
SCR-Net \cite{li2022structure} employed structure consistency to cultivate an IQE model for real cataract data from synthesized cataract datasets.
PCE-Net \cite{liu2022degradation} utilized a feature pyramid in IQE to bolster resilience against quality degradations in image embedding.
Furthermore, GFE-Net \cite{Agenericfundusimageenhancementnetwork} devised a frequency self-supervision pretext to robustly reconstruct structure and high-quality images from images afflicted by various degradations.
       
% As for fundus IQE, \cite{guo2024multi} point out that existing fundus IQE strategies only target at specific type of degradations. However, in real clinical scenarios, degradation can vary in levels and types. In some cases, low-level and additive artifacts exist in fundus images, while in other domains, degradations are multiplicative and high-level blurring. Consequently, \cite{guo2024multi} proposes a dynamic filtering network based on U-Net to adaptively detect distinct degradation and generate effective filters accordingly. 

\subsubsection{Interpretability}
Learning-based methods facilitate the advancements in IQA and IQE algorithms.
However, the majority of these approaches primarily emphasize overall performance, overlooking the need for interpretable feedback.
These learning-based models frequently operate as "black boxes," obscuring their internal decision-making processes. This lack of transparency can erode trust, particularly in medical scenarios, where understanding why a model makes a specific judgment is crucial. Furthermore, when training data is limited, interpretability becomes even more vital for ensuring accountability and building confidence in the reliability of the model's predictions.

\begin{figure}[!ht]
    \begin{centering}
        \includegraphics[width=\linewidth]{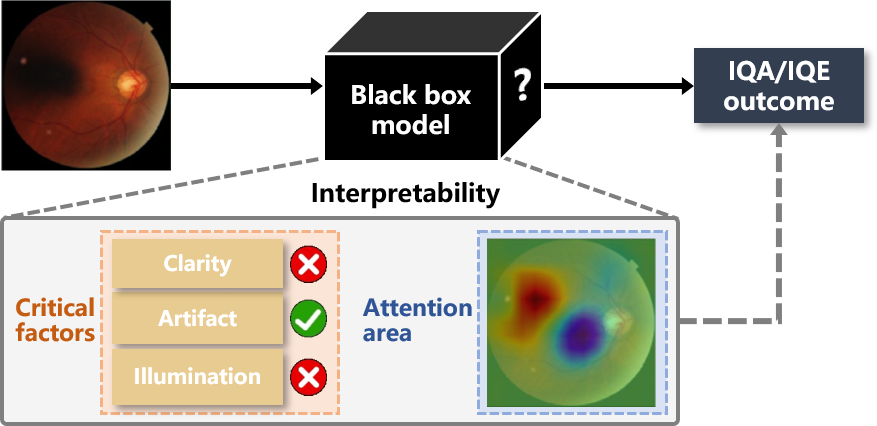}
        \par
    \end{centering}
\caption{Illustration of ways to improve interpretability of IQA/IQE models.}
\vskip -5pt
\label{fig:application_interpretability}
\end{figure}

Interpretable fundus IQA and IQE algorithms assist ophthalmologists in comprehending the usability of fundus images and the reliability of enhancements.
As depicted in Fig.~\ref{fig:application_interpretability}, current research underscores the interpretability of fundus IQA and IQE models through two main paradigms: 1) Exploring critical factors to provide interpretable outcomes, and 2) Producing attention feedback to elucidate the models' decision-making processes.

In the process of fundus IQA, ophthalmologists commonly take into account factors like image clarity, presence of artifacts, and lighting conditions, which are integrated to determine an overall quality grade. Accordingly, achieving interpretability in fundus IQA and IQE involves predicting these critical factors individually along with the final overall quality grade.
Fu et al.~\cite{fu2019evaluation} introduced a multi-color spatial fusion network to amalgamate multi-source features for a comprehensive evaluation of fundus image quality.
Shen et al. \cite{shen2020domain} deconstructed fundus IQA into elements of artifacts, clarity, and field definition, utilizing neural networks to assign weights to each factor and provide an interpretable overall quality assessment.
Li et al. \cite{li2023deepquality} introduced an image quality comprehensive score that considers the illumination, clarity, and integrity of fundus images, as well as created heatmaps to identify visually the specific areas within an image that contribute to quality issues.
Yi et al. \cite{yi2023label} proposed a method that merges the zero-shot classification results of Contrastive Language-Image Pre-Training (CLIP)~\cite{radford2021learning} with frequency and time domain features to create a semantics-aware contrastive learning model for fundus IQA.

Attention feedback represents an alternative paradigm to interpret the decision-making of models. Techniques such as Class Activation Mapping (CAM) enable the highlighting of fundus regions that the model prioritizes, facilitating the reassessment of outcomes and understanding of the decision rationale. Khalid et al. \cite{khalid2024fgr} employed Gradient and CAM to visualize attention feedback, delivering insights into the decision-making process.        
     
\subsection{Fundus image understanding boosted by IQA and IQE}
Beyond the progression of fundus IQE and IQA algorithms, researchers have explored how these techniques can boost downstream fundus image analysis. The promotion stemming from IQE and IQA for fundus image analysis manifests in two primary ways: 1)  Leveraging IQA and IQE methods to either filter out unqualified images or enhance their quality, guaranteeing high-quality inputs for downstream tasks. 2) Incorporating features from fundus IQA and IQE as supplementary data to contribute to fundus image understanding.

\subsubsection{Quality guarantee for downstream tasks}
\begin{figure}[!ht]
    \begin{centering}
        \includegraphics[width=\linewidth]{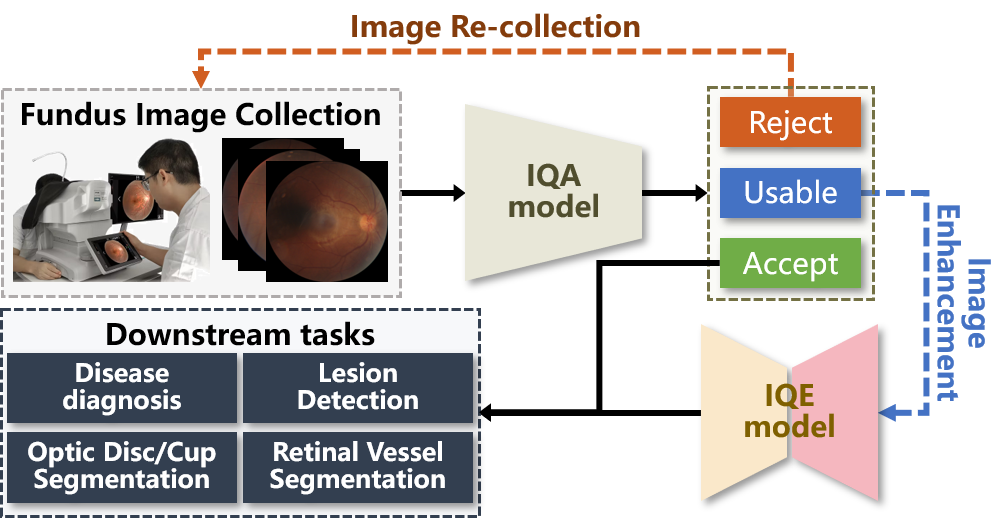}
        \par
    \end{centering}
\caption{Integrated workflow of fundus IQA and IQE into downstream tasks to guarantee image quality.}
\vskip -5pt
\label{fig:application_quality_guarantee}
\end{figure}

An immediate application of IQA and IQE methods is to guarantee the quality and visibility of collected fundus images for downstream tasks, such as disease diagnosis and lesion detection.
As illustrated in Fig.~\ref{fig:application_quality_guarantee}, the collected fundus images undergo an initial assessment by the IQA model. High-quality samples are accepted and passed on directly to downstream tasks, while low-quality samples that cannot be properly enhanced are rejected and necessitate recollection. Usable samples whose degradation can be restored through IQE, are enhanced and forwarded to downstream tasks.

\noindent\textbf{IQA Application}

As summarized in Table~\ref{tab:datasets}, most previous datasets and research categorize fundus images into two quality grades (high or low).
Nevertheless, in clinical scenarios, certain low-quality images may lack crucial details that are beyond the scope of IQE for resolution, while others are merely degraded and can be substantially enhanced through IQE. Consequently, for effective support of downstream tasks, it becomes imperative to differentiate low-quality images based on their amenability to enhancement~\cite{fu2019evaluation}.

In the diagnosis for cataract patients, Pratap et al. \cite{pratap2019computer} utilized the IQA metrics NIQE and PIQE to select training and testing data, thereby enhancing the training and predictive capabilities of deep neural networks. Additionally, IQA techniques were leveraged in \cite{yadav2022computer} to aid in the diagnosis of cataracts through support vector machines and ensemble learning.
For boosting clinical diagnosis, Liu et al. \cite{liu2023deepfundus} selected fundus images through a comprehensive assessment of the position, illumination, and clarity of various fundus organs.
Li et al. \cite{li2023deepquality} introduced fundus IQA to facilitate the enhancement process for producing higher-quality images, subsequently improving the diagnosis of retinopathy of prematurity.
    
\noindent\textbf{IQE Application}

Enhancing imaging quality through IQE boosts fundus observation, thereby advancing clinical analysis and diagnosis.
% Some other fundus images possess part of degradation. However, they still contain rich information, and re-collection lowers the clinical efficiency and patients' experiences. Under such circumstances, some methods introduce IQE instead to improve the quality of the image.

In fundus diagnosis, Kaushik et al. \cite{kaushik2021diabetic} standardized the intensity of the red, green, and blue channels to eliminate the effects of uneven illumination distribution in DR diagnosis. 
Similarly, Adal et al. \cite{adal2017automated} normalized the distribution of background brightness and contrast for all captured fundus images to facilitate early abnormal detection and screening for DR.
Wu et al. \cite{wu2020attennet} utilized Otsu's binarization, morphological opening and closing operations, and the Sobel operator to improve image quality, leading to more precise fundus diagnosis.
To facilitate DR grading, Hou et al. \cite{hou2022image} proposed a collaborative learning framework that simultaneously trains the subnetworks responsible for image quality assessment, enhancement, and DR disease grading within a unified framework.
Through domain adaptation bridging synthetic and authentic cataract data, Li et al. \cite{li2022annotation} facilitated the diagnosis of multiple ocular diseases in fundus images of cataract patients.
Through simultaneously conducting IQE and domain adaptation in fundus images, Guo et al. \cite{Bridging} improved ocular disease recognition in the ODIR datasets.

In fundus segmentation, Shyamalee et al. \cite{shyamalee2022glaucoma} applied a median filter to remove noise from the fundus images and then used CLAHE to enhance image quality and contrast to boost fundus vessel segmentation and glaucoma diagnosis.
Shanthamalar et al. \cite{shanthamalar2022automatic} applied IQE to minimize the distinctions between bright and dark fundus areas, thereby enhancing segmentation precision.
The generalizable IQE models developed by \cite{li2022structure} and \cite{Agenericfundusimageenhancementnetwork} provided enhancement invariant degradation variance and boosted the fundus structure segmentation in images with cataracts and degradations, respectively.
Kaur et al. \cite{kaur2023detection} merged conventional computer vision with deep learning to enhance image quality for improved fundus vessel segmentation. 
The source-free domain adaptive IQE algorithm presented in \cite{EnhancingandAdapting}, allows facilitating fundus vessel segmentation using post-deployment data.

\subsubsection{Quality feature fusion for downstream tasks}
\begin{figure}[htbp]
    \begin{centering}
        \includegraphics[width=\linewidth]{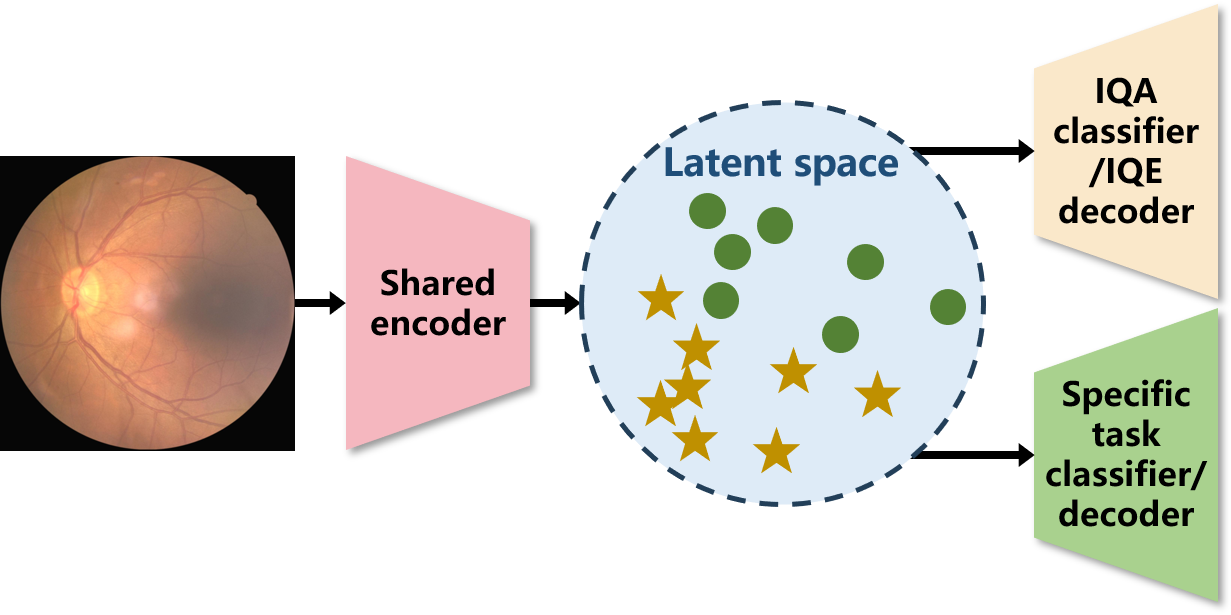}
        \par
    \end{centering}
\caption{Framework for integrating fundus IQA and IQE with downstream tasks.}
\vskip -5pt
\label{fig:application_feature_fusion}
\end{figure}

Instead of directly applying IQA and IQE to guarantee image quality in inputs for downstream tasks, an alternative paradigm has been proposed aiming to forge deeper connections between quality factors and medical features in latent spaces as illustrated in Fig.~\ref{fig:application_feature_fusion}.

Approaches within this paradigm typically adopt a multi-task learning framework, involving a branch dedicated to IQA or IQE alongside another branch for downstream tasks.
As shown in Fig.~\ref{fig:application_feature_fusion}, a captured fundus image undergoes embedding via a shared encoder, introducing quality factors and medical features into a fused identical latent space.
The integration of features through implicit co-embedding, as suggested within the multi-task learning setup, could be viewed as an inductive transfer, thereby enhancing the overall performance of each task.

Dai et al. \cite{dai2021deep} developed a DR grading system that employed deep neural networks to predict image quality issues and shared parameters with DR grading, facilitating a more precise assessment of disease severity.
Hou et al. \cite{hou2022image} presented a dual-path strategy for IQA and segmentation, which identifies degradation factors in low-quality images and utilizes multitask learning to enhance segmentation accuracy.
Che et al. \cite{che2023image} explicitly added a supplementary branch to merge diagnostic and quality assessment features, guiding the model to identify the optimal mapping of feature spaces, thereby enhancing both diagnosis and IQA performance.

\subsection{Challenges and Insight}
While fundus IQA and IQE have achieved notable successes in laboratories, their deployment in clinical scenarios remains a frontier for exploration. Addressing challenges such as data scarcity, annotation limitations, model generalizability, and outcome interpretability are critical hurdles impeding the clinical implementation of these models. 
Furthermore, realizing the full potential of IQA and IQE for improving clinical analysis and diagnosis requires further development.
Therefore, encouraging the integration of IQA and IQE algorithms in clinical problem-solving holds significant promise, with substantial promise for further advancements and applications in practice.

\section{Future Outlook}
Despite the substantial advancements achieved in algorithms and performance of fundus IQA and IQE, there are still barriers to overcome for fully leveraging their capabilities in clinical scenarios. 
According to clinical demands and the current development of IQA and IQE, we discuss potential future research directions.

\subsubsection{Medical knowledge embedding}
Generic IQA and IQE methods may prioritize visually appealing aspects over medically relevant ones, necessitating the integration of medical expertise to emphasize the perception of diagnostically crucial features.
Previous studies have integrated medical knowledge, such as anatomical structures~\cite{xu2023deep} \cite{li2022annotation}, to assess and enhance fundus images.
Nevertheless, there remains substantial potential for further exploration in effectively embedding medical expertise into IQA and IQE algorithms. 
Encouragingly, recent advancements in foundational models have introduced novel embedding strategies for human knowledge.
For instance, Yi et al. \cite{yi2023label} utilized the knowledge embedded in CLIP to develop a semantics-aware model for fundus IQA. Furthermore, Liang et al. \cite{liang2023iterative} introduced open-world knowledge from CLIP to boost enhancement networks.
Therefore, leveraging foundation models holds substantial promise for advancing medical knowledge embedding in fundus IQA and IQE.

\subsubsection{Interference interpretation}
Present IQA and IQE techniques have facilitated the identification of qualified fundus images~\cite{li2023deepquality} and the enhancement of visual quality~\cite{liu2023deepfundus}. However, there remains a deficiency in interpreting interference to support both image acquisition and enhancement processes.
Interpreting interference has proven beneficial in low-light image enhancement, leading to simplified models and improved performance~\cite{wang2024extracting}.
Interference interpretation can reveal valuable insights into the image formation mechanisms and potential anomalies within the acquisition system, which are crucial for guiding clinical imaging procedures and facilitating reliable image enhancement.
While identifying explicit degradations helps achieve interpretable assessment and targeted enhancement in fundus images, there exists a gap between the identified degradation and the interferences causing such degradation.
Consequently, progress in interference interpretation will further facilitate the clinical applicability and effectiveness of these techniques.

\subsubsection{Information decouple}
Fundus images can be broadly categorized into visual information (e.g., color, brightness) and semantic information (e.g., anatomical structures, abnormal objects). This distinction allows fundus IQA and IQE to better align with ophthalmologists' perceptions. For instance, a fundus image with minor visual imperfections might be acceptable if the semantic information is clear, whereas a visually flawless image with distorted semantic content would be deemed unsuitable. While semantic information has been independently used in fundus IQA and IQE~\cite{Agenericfundusimageenhancementnetwork}, this decomposition has traditionally relied on manual annotation. Notably, recent advancements in semantics-aware modules for single image restoration~\cite{wu2024towards} and super-resolution~\cite{wu2024seesr} are promoting automated decomposition of visual and semantic information in image processing.

\subsubsection{Dynamic outcome personalization}
Variations in imaging equipment and physician reading habits introduce visual discrepancies in fundus images across different clinical centers. These discrepancies can lead to inconsistencies between the output of pre-trained IQA and IQE models and clinicians' expectations. Consequently, providing personalized model outcomes according to the requirements of each clinical center is crucial for improving their practical value in clinical diagnosis and treatment.
However, previous studies have been limited to static model parameters, requiring additional training to adjust these outcomes.
The recent advancement in prompt engineering enables dynamic generation of personalized outputs, such as specific color~\cite{kosugi2024prompt} and image style~\cite{lee2025diffusion}.
This dynamic personalization is expected to further advance the clinical application of fundus IQA and IQE.

\subsubsection{Continual optimization}
Current intelligent algorithms for medical images primarily rely on pre-trained static models, which are challenging to optimize further in real-world clinical scenarios. Regional variations in imaging equipment and demographic differences in patient populations can introduce biases in these static models when applied to new data. 
However, the abundance of data generated in clinical practice represents a valuable resource for enhancing model performance. Thus continuous optimization, incorporating clinical test data, offers a cost-effective way to mitigate these biases and steadily improve model performance without requiring additional data collection~\cite{pianykh2020continuous}. While some algorithms have successfully optimized fundus IQE models using clinical test data through techniques like source-free domain adaptation~\cite{EnhancingandAdapting}, and continuous learning has been explored for image enhancement in general~\cite{jung2024clode}, an open area remains in applying continuous learning specifically to medical image enhancement.

\subsubsection{Cross-modality generalization}
With advancements in imaging theory and techniques, alternative imaging modalities to traditional fundus photography, such as scanning laser ophthalmoscope (SLO) and ultra-wide-field (UWF) scanning, are gaining popularity.
As these emerging modalities still encounter challenges involving image quality in clinical scenarios, IQA and IQE algorithms remain meaningful.
However, the availability of data and annotations for these modalities is currently more limited compared to the traditional one.
Researchers have attempted to generalize existing intelligent models based on traditional data and annotations to these emerging modalities, reducing the cost of model development. 
Despite the success of generalizing segmentation~\cite{li2025aif} and diagnosis models~\cite{ran2024source}, further efforts are required to implement cross-modality generalization in image quality issues.
This is because IQA and IQE tasks are intricately coupled to specific imaging systems~\cite{shen2020modeling}, presenting additional obstacles in generalizing models across diverse modalities.

\section{Conclusion}
In clinical scenarios, fundus photography often experiences quality degradation, which can significantly hamper the subsequent diagnosis and treatments. Consequently, fundus IQA and IQE are essential to ensure the clinical utility of these images.
This paper presents a comprehensive review of fundus IQA and IQE algorithms, recent research advances, and practical applications.
We begin by outlining the fundamentals of fundus photography and common image artifacts, and then systematically summarize the different paradigms in fundus IQA and IQE.
Moreover, we discuss the practical challenges and solutions for deploying fundus IQA and IQE in clinics and offer insights into future research directions to advance clinical examination by fundus photography.

\bibliographystyle{IEEEtran}
\bibliography{refs}

\end{document}